\pdfoutput=1
\documentclass[11pt,aps,a4paper,eqsecnum,amsmath,amssymb
              ,nofootinbib
              ,notitlepage,endfloats*]{revtex4-1}

\usepackage{graphicx}

\def\notiz[#1]{\textbf{
#1}}

\def\Wop[#1][#2][#3][#4][#5]{W
  \left(\left.\begin{array}{cc}
        #1&#2\\#3&#4\end{array}\right|\,#5
  \right)}
\def\Bop[#1][#2][#3][#4]{B
  \left(\left.\begin{array}{c}
        #1\\#3\end{array}\,#2\,\right|\,#4
  \right)}
  \def\BopP[#1][#2][#3][#4]{B'
  \left(\left.\begin{array}{c}
        #1\\#3\end{array}\,#2\,\right|\,#4
  \right)}
\def\BopR[#1][#2][#3][#4]{B
  \left(\left. #2 \, \begin{array}{c}
       #1\\#3\end{array}\,\right|\,#4
  \right)}
  \def\BopRP[#1][#2][#3][#4]{B'
  \left(\left. #2 \, \begin{array}{c}
       #1\\#3\end{array}\,\right|\,#4
  \right)}
\def\WopB[#1][#2][#3][#4][#5]{\bar{W}
  \left(\left.\begin{array}{cc}
        #1&#2\\#3&#4\end{array}\right|\,#5
  \right)}

\begin{document}
 \preprint{}
\title{Inversion identities for inhomogeneous face models}   

\author{Holger Frahm}
\author{Nikos Karaiskos}
\affiliation{%
Institut f\"ur Theoretische Physik, Leibniz Universit\"at Hannover,
Appelstra\ss{}e 2, 30167 Hannover, Germany}

\date{\today}

\begin{abstract}
  We derive exact inversion identities satisfied by the transfer matrix of
  inhomogeneous interaction-round-a-face (IRF) models with arbitrary boundary
  conditions using the underlying integrable structure and crossing properties
  of the local Boltzmann weights.  For the critical restricted solid-on-solid
  (RSOS) models these identities together with some information on the
  analytical properties of the transfer matrix determine the spectrum
  completely and allow to derive the Bethe equations for both periodic and
  general open boundary conditions.
\end{abstract}

\maketitle

\section{Introduction}
Functional relations between the transfer matrices of integrable models
together with the knowledge of their analytical properties provide a powerful
basis for the solution of their spectral problem.  An important example are the 
so called inversion relations \cite{Resh83,Pear87a,Pear87b}.  In the
thermodynamic limit these relations become identities (at least for part of
the spectrum) allowing to compute the free energy of certain models exactly
\cite{Stro79,Schu81}.  Generalized inversion relations for the restricted
solid-on-solid (RSOS) model have been obtained from the fusion hierarchy
\cite{BaRe89,BePO96} and have been used to identify the low energy effective
theory of the critical model through solution of nonlinear integral equations
\cite{KlPe92} or to study their surface critical behaviour \cite{ZhBA96}.

Recently, sets of \emph{exact} inversion identities for the transfer matrices
of inhomogeneous vertex models have been used to tackle the long-standing
problem of finding Bethe equations for the spectrum of the integrable XXZ spin
chain subject to non-diagonal boundary conditions which break the $U(1)$
symmetry of the bulk, see e.g.\ Refs.\
\cite{Nepo02,Nepo03,Nepo04,CaoX03,BaKo07,Galleas08,
  FrSW08,FGSW11,Nicc12,CrRa12}.  They have been derived for vertex models
using as only input the underlying Yang-Baxter and reflection equation and
physical assumptions such as crossing and unitary of the local Boltzmann
weights.  Unlike the inversion identities mentioned above they only hold for a
discrete set of spectral parameters related to the inhomogeneities introduced
in the lattice model \cite{CYSW13a}.  Similar expressions for the
corresponding eigenvalues had been obtained before using Sklyanin's separation
of variables \cite{Skly92,FrSW08}, or by considering certain matrix elements of
the transfer matrix \cite{{CaoX13,CYSW13}}.

While the use of these identities for the actual computation of eigenvalues is
restricted to small systems they allow, once complemented by information on
the analytical properties of the transfer matrix, to formulate the spectral
problem in the form of Baxter's $TQ$-equation \cite{Baxter:book} or
inhomogeneous generalisations thereof \cite{CaoX13,CYSW13,CYSW13a}.
In addition, the number of solutions to the inversion identities are rather
easily counted which allows to address the problem of completeness of the
Bethe ansatz for the underlying model
\cite{Skly92,FrSW08,Nepo13,FaKN14,KiMN14}.

First attempts to extend this method for the solution of the spectral problem
to integrable interaction-round-a-face (IRF) statistical models have made use
of the reorganisation of Boltzmann weights of solid-on-solid (SOS) models in
an $R$-matrix solving the dynamical six-vertex Yang-Baxter algebra
\cite{FeVa96a}.  Adapting Sklyanin's separation of variables the eigenvalues of
the transfer matrix of the dynamical six-vertex model on a lattice with odd
number of sites and with antiperiodically twisted boundary conditions have
been shown to satisfy quadratic equations for a discrete set of spectral
parameters \cite{Nicc13}.  In another approach functional relations have been
derived from the dynamical Yang-Baxter equation to determine the partition
function of the SOS model with domain wall boundaries \cite{Gall12,Gall13}.

In this paper we derive inversion identities for the transfer matrix of
general IRF models directly in the face formulation of the Yang-Baxter algebra
using unitarity and crossing properties of the local Boltzmann weights.  For
this we consider inhomogeneous face models subject to periodic and generic
integrable open boundary conditions.  Starting from these identities we show
that they allow to derive $TQ$-equations for the critical RSOS models.  The
eigenvalues of the transfer matrix are parametrized in terms of the solution
to Bethe equations which allow to study properties of finite chains and to
perform the thermodynamic limit.

\section{Discrete inversion identities for IRF models}

Below we consider inhomogeneous IRF models and construct inversion identities
satisfied by their commuting transfer matrices.  As will become transparent
below, the derivation is valid for a generic class of integrable lattice
models, provided that certain local relations are satisfied. The fundamental
blocks of the models are given by the Boltzmann face weights

\begin{picture}(150,50)(0,-5)
\put(130,11){$ \Wop[a][b][d][c][u]$~~~~~= }
\put(250,0){\line(0,1){30}}
\put(250,0){\line(1,0){30}}
\put(280,0){\line(0,1){30}}
\put(250,30){\line(1,0){30}}
\put(262,12){$u$}
\put(241,33){$a$}
\put(282,33){$b$}
\put(241,-9){$d$}
\put(282,-9){$c$}
\end{picture}

\noindent
where the spin variables $a,b,c,d$ take values within a discrete set
$\mathfrak{S}$.  The allowed states of the IRF model are constrained by
selection rules which are conveniently encoded in the so-called adjacency
matrix $A$:
\begin{equation}
A_{ab} = \left\{ 
\begin{array}{c}
  0\, : ~~~ \textrm{spins $a$ and $b$ may not be adjacent} \cr 
  1\, : ~~~ \textrm{spins $a$ and $b$ may be adjacent,~~~~}  
\end{array}
\right. 
\nonumber
\end{equation}
such that the Boltzmann weights satisfy
\begin{equation}
 \Wop[a][b][d][c][u] = A_{ab} \, A_{bc} \, A_{cd} \, A_{da} \,
 \Wop[a][b][d][c][u] \, . 
\end{equation}

The face weights are assumed to satisfy a set of local relations.  First of
all, the integrability of the models is guaranteed by the Yang-Baxter equation
(YBE)
\begin{equation}
\begin{split}
& \sum_{g} \Wop[f][g][a][b][u-v] \Wop[g][d][b][c][u] \Wop[f][e][g][d][v] = \cr
& \qquad\qquad \sum_{g} \Wop[a][g][b][c][v] \Wop[f][e][a][g][u] 
\Wop[e][d][g][c][u-v]\, .
\end{split}
\label{YBE}
\end{equation}
In addition, we assume that the Boltzmann weights satisfy unitarity
\begin{equation}
\sum_{e} \Wop[d][e][a][b][u] \Wop[d][c][e][b][-u] = \rho(u) \, \rho(-u) \,
  \delta_{ac} \, ,
\label{W_Unitarity} 
\end{equation}
crossing symmetry
\begin{equation}
\Wop[b][c][a][d][\lambda - u] = \Wop[a][b][d][c][u] \, ,
\label{Wcrossym} 
\end{equation}
and become diagonal at the so-called shift points 
\begin{equation}
\Wop[a][b][d][c][0] =  \delta_d^b \, , \qquad \textrm{and}  \qquad
\Wop[a][b][d][c][\lambda] =  \delta^a_c\, ,
\label{Wvalues}
\end{equation}
which correspond to the identification of scattering particles in the
underlying physical picture. The function $\rho(u)$ appearing in
(\ref{W_Unitarity}) is model-dependent.  It can be normalized such that
$\rho(0)=1$.

For models with open boundary conditions (left and right) boundary Boltzmann
weights have to be introduced
\begin{equation*}
\begin{picture}(100,50)(50,-10)
\put(30,11){$ \Bop[a][b][c][u]$~~~~~= }
\put(140,-5){\line(0,1){40}}
\put(140,-5){\line(1,1){20}}
\put(140,35){\line(1,-1){20}}
\put(137,38){$a$}
\put(163,10){$b$}
\put(137,-13){$c$}
\put(144,12){$u$}
\end{picture}
\end{equation*}
Integrability requires that they satisfy the
reflection or boundary Yang-Baxter equation (BYBE).  For the left boundary
weights the BYBE is given by \cite{Kulish96,BePO96}
\begin{equation}
\begin{split}
 & \sum_{f,g} \Wop[c][b][f][a][u-v] \Wop[d][c][g][f][\lambda - u - v] 
 \Bop[g][f][a][u] \Bop[e][d][g][v] = \cr
 & \qquad \qquad \sum_{f,g} \Wop[e][d][f][c][u-v] 
 \Wop[f][c][g][b][\lambda-u-v] \Bop[e][f][g][u] \Bop[g][b][a][v] \, .
\end{split}
\label{BYBE}
\end{equation}
The boundary weights are normalized by the boundary inversion condition
\begin{equation}
  \sum_{c} \Bop[a][b][c][u] \Bop[c][b][d][-u] = \beta_a(u) 
  \, \beta_a(-u)\,\delta_d^a \,  , 
\label{bound_inv}
\end{equation}
with model-dependent functions $\beta_a(u)$.  Furthermore, they are required
to satisfy the boundary crossing condition
\begin{equation}
  \label{bcross}
  \sum_{d} \Bop[c][d][a][u]\,\Wop[c][b][d][a][2u-\lambda]
  = -\rho(\lambda-2u)\, \Bop[c][b][a][\lambda-u]\, .
\end{equation}
Similar relations hold for the right boundary weights \cite{BePO96}.

For the derivation of inversion identities below we use a graphical
representation of the relations listed above, see also \cite{BePO96}: the YBE
(\ref{YBE}) is given by

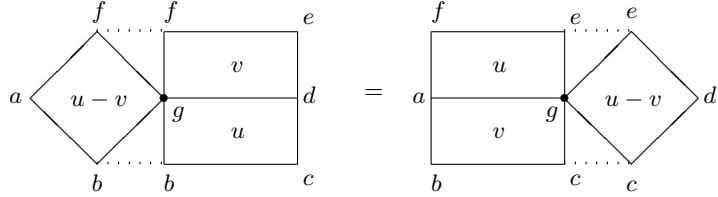
\begin{picture}(400,80)(0,-10)
\footnotesize
\put(100,0){\line(-1,1){25}}
\put(100,0){\line(1,1){25}}
\put(75,25){\line(1,1){25}}
\put(100,50){\line(1,-1){25}}
\put(90,22){$u-v$}
\put(67,23){$a$}
\put(98,55){$f$}
\put(98,-10){$b$}

\put(125,25){\circle*{3}}
\put(125,25){\line(0,1){25}}
\put(125,25){\line(0,-1){25}}
\put(125,50){\line(1,0){50}}
\put(125,25){\line(1,0){50}}
\put(125,0){\line(1,0){50}}
\put(175,25){\line(0,1){25}}
\put(175,25){\line(0,-1){25}}
\put(150,35){$v$}
\put(150,10){$u$}

\multiput(100,0)(4,0){7}{\line(0,1){1}}
\multiput(100,50)(4,0){7}{\line(0,1){1}}

\put(128,18){$g$}

\put(125,55){$f$}
\put(125,-10){$b$}
\put(177,53){$e$}
\put(177,23){$d$}
\put(177,-8){$c$}

\put(200,25){$=$}

\put(225,25){\line(0,1){25}}
\put(225,25){\line(0,-1){25}}
\put(225,50){\line(1,0){50}}
\put(225,25){\line(1,0){50}}
\put(225,0){\line(1,0){50}}
\put(275,25){\line(0,1){25}}
\put(275,25){\line(0,-1){25}}
\put(248,35){$u$}
\put(248,10){$v$}

\put(225,55){$f$}
\put(225,-10){$b$}
\put(277,53){$e$}
\put(277,-8){$c$}
\put(218,23){$a$}

\put(300,0){\line(-1,1){25}}
\put(300,0){\line(1,1){25}}
\put(275,25){\line(1,1){25}}
\put(300,50){\line(1,-1){25}}
\put(275,25){\circle*{3}}
\put(290,22){$u-v$}
\put(268,18){$g$}
\put(298,55){$e$}
\put(298,-10){$c$}
\put(327,23){$d$}

\multiput(275,0)(4,0){7}{\line(0,1){1}}
\multiput(275,50)(4,0){7}{\line(0,1){1}}

\end{picture}

\noindent 
Here and in the following diagrams the spin variables on nodes with a solid
circle are summed over all elements from $\mathfrak{S}$.  Nodes with equal
spins are connected by a dotted line.  Similarly, the unitarity condition is
represented by

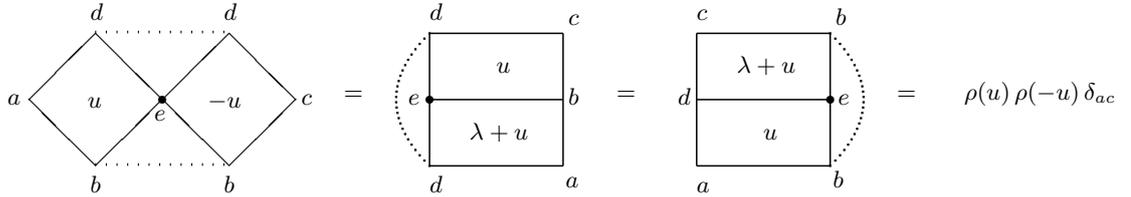
\begin{picture}(400,80)(0,-10)
\footnotesize
\put(50,0){\line(-1,1){25}}
\put(50,0){\line(1,1){25}}
\put(25,25){\line(1,1){25}}
\put(50,50){\line(1,-1){25}}
\put(47,22){$u$}
\put(17,23){$a$}
\put(48,55){$d$}
\put(48,-10){$b$}

\multiput(50,0)(4,0){13}{\line(0,1){1}}
\multiput(50,50)(4,0){13}{\line(0,1){1}}

\put(75,25){\circle*{3}}

\put(100,0){\line(-1,1){25}}
\put(100,0){\line(1,1){25}}
\put(75,25){\line(1,1){25}}
\put(100,50){\line(1,-1){25}}
\put(92,22){$-u$}
\put(72,17){$e$}
\put(98,55){$d$}
\put(98,-10){$b$}
\put(127,23){$c$}

\put(143,25){$=$}

\linethickness{0.3mm}

\qbezier[25](175,50,0)(150,25)(175,0)

\linethickness{0.2mm}

\put(175,25){\circle*{3}}
\put(175,25){\line(0,1){25}}
\put(175,25){\line(0,-1){25}}
\put(175,50){\line(1,0){50}}
\put(175,25){\line(1,0){50}}
\put(175,0){\line(1,0){50}}
\put(225,25){\line(0,1){25}}
\put(225,25){\line(0,-1){25}}
\put(200,35){$u$}
\put(190,10){$\lambda + u$}
 
\put(167,23){$e$}

\put(175,55){$d$}
\put(175,-10){$d$}
\put(227,53){$c$}
\put(227,23){$b$}
\put(226,-8){$a$}

\put(245,25){$=$}

\put(325,25){\circle*{3}}
\put(275,25){\line(0,1){25}}
\put(275,25){\line(0,-1){25}}
\put(275,50){\line(1,0){50}}
\put(275,25){\line(1,0){50}}
\put(275,0){\line(1,0){50}}
\put(325,25){\line(0,1){25}}
\put(325,25){\line(0,-1){25}}
\put(290,35){$\lambda + u$}
\put(300,10){$u$}
 
\put(268,23){$d$}

\linethickness{0.3mm}

\qbezier[25](325,50,0)(350,25)(325,0)

\put(275,55){$c$}
\put(275,-10){$a$}
\put(327,53){$b$}
\put(328,23){$e$}
\put(326,-8){$b$}

\put(350,25){$=~~~~~ 
\rho(u) \, \rho(-u) \, \delta_{ac} $} 

\end{picture}

\noindent
where the first diagram depicts (\ref{W_Unitarity}).  Crossing symmetry
(\ref{Wcrossym}) has been used for the alternative representations.

Boundary inversion (\ref{bound_inv}) and crossing condition (\ref{bcross}) 
for the left boundary weights are represented in a similar manner by

\noindent
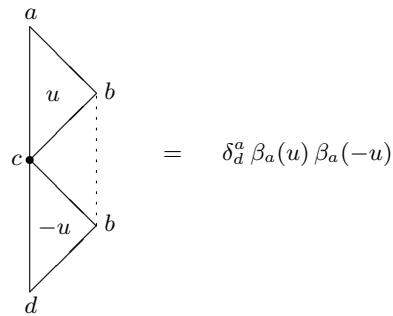
\begin{picture}(400,130)(-150,-10)
\footnotesize

\put(0,0){\line(0,1){100}}
\put(0,0){\line(1,1){25}}
\put(0,50){\line(1,-1){25}}
\put(0,100){\line(1,-1){25}}
\put(0,50){\line(1,1){25}}
\put(0,50){\circle*{3}}

\multiput(25,25)(0,4){13}{\line(0,1){1}}

\put(-2,-8){$d$}
\put(-2,103){$a$}
\put(-7,48){$c$}
\put(28,23){$b$}
\put(28,73){$b$}

\put(3,22){$-u$}
\put(6,72){$u$}

\put(50,50){$=~~~~ \delta_d^a \, \beta_a(u) 
\, \beta_a(-u)$}

\end{picture}

\noindent and

\begin{picture}(400,80)(-100,-10)
\footnotesize

\put(0,0){\line(0,1){50}}
\put(0,0){\line(1,1){50}}
\put(0,50){\line(1,-1){50}}
\put(50,50){\line(1,-1){25}}
\put(50,0){\line(1,1){25}}

\multiput(0,0)(4,0){13}{\line(0,1){1}}
\multiput(0,50)(4,0){13}{\line(0,1){1}}

\put(25,25){\circle*{3}}

\put(7,23){$u$}
\put(36,23){$2u-\lambda$}

\put(-2,-7){$a$}
\put(48,-7){$a$}
\put(-2,53){$c$}
\put(48,53){$c$}
\put(77,23){$b$}
\put(22,15){$d$}


\put(100,25){$=~~~ -\rho(\lambda-2u)$}

\put(175,0){\line(0,1){50}}
\put(175,0){\line(1,1){25}}
\put(175,50){\line(1,-1){25}}

\put(175,23){$\lambda - u$}
\put(173,-7){$a$}
\put(173,53){$c$}
\put(203,23){$b$}

\end{picture}

\noindent
respectively.  The corresponding relations for the right
boundaries are obtained by reflecting these diagrams.

\subsection{Periodic boundary conditions}
To derive a set of inversion identities for integrable IRF models subject to
periodic boundary conditions we introduce columns of inhomogeneities
$\{u_\ell\}$.  The resulting transfer matrix is given by the product of
Boltzmann weights
\begin{equation}
\label{trnsf_mtx_period}
\begin{aligned}
  \mathbf{T}(u) &\equiv 
  T_{b_{0} \cdots b_{L}}^{a_{0} \cdots a_{L}} (u) = \prod_{\ell = 1}^L 
  \Wop[a_{\ell-1}][a_\ell][b_{\ell-1}][b_{\ell}][u-u_\ell] \, ,
  \\
  &\begin{picture}(350,60)(-30,-10)

    \put(-25,11){$=$}

    \footnotesize
    \put(0,0){\line(1,0){300}}
    \put(0,0){\line(0,1){25}}
    \put(0,25){\line(1,0){300}}
    \put(300,0){\line(0,1){25}}
    \put(50,0){\line(0,1){25}}
    \put(250,0){\line(0,1){25}}
    \put(125,0){\line(0,1){25}}
    \put(175,0){\line(0,1){25}}

    \put(13,10){$u-u_1$}
    \put(83,10){$\cdots$}
    \put(138,10){$u-u_k$}
    \put(207,10){$\cdots$}
    \put(263,10){$u-u_L$}

    \put(-2,-10){$b_0$}
    \put(47,-10){$b_1$}
    \put(297,-10){$b_L$}
    \put(247,-10){$b_{L-1}$}
    \put(122,-10){$b_{k-1}$}
    \put(172,-10){$b_k$}
    \put(-2,30){$a_0$}
    \put(47,30){$a_1$}
    \put(122,30){$a_{k-1}$}
    \put(172,30){$a_k$}
    \put(247,30){$a_{L-1}$}
    \put(297,30){$a_L$}

  \end{picture}
\end{aligned}
\end{equation}
where $(a_{L}, b_{L})\equiv(a_{0}, b_{0})$ to impose periodic boundary
conditions.  As a direct consequence of the YBE (\ref{YBE}) the transfer
matrices form a commuting family of operators,
$[\mathbf{T}(u),\mathbf{T}(v)]=0$, which establishes the integrability of the
inhomogeneous model.
Using the local relations (\ref{W_Unitarity})-(\ref{Wvalues}) then, it can be
shown that the product $\mathbf{T}(u) \, \mathbf{T}(\lambda + u)$ becomes
diagonal for $u=u_k$. Using the graphical representation introduced above, we
obtain
\begin{equation*}
\begin{aligned}
 & T_{b_{0} \cdots b_{L}}^{a_{0} \cdots a_{L}} (u_k) \,
   T_{c_{0} \cdots c_{L}}^{b_{0} \cdots b_{L}} (\lambda + u_k)  = 
 \\
& \begin{picture}(450,80)(-27,-20)

\put(-25,24){$=$}
\footnotesize

\put(0,0){\line(1,0){300}}
\put(0,0){\line(0,1){50}}
\put(0,25){\line(1,0){300}}
\put(300,0){\line(0,1){50}}
\put(50,0){\line(0,1){50}}
\put(250,0){\line(0,1){50}}
\put(125,0){\line(0,1){50}}
\put(175,0){\line(0,1){50}}
\put(0,50){\line(1,0){300}}

\put(13,35){$u_k-u_1$}
\put(83,35){$\cdots$}
\put(138,35){$u_k-u_k$}
\put(207,35){$\cdots$}
\put(263,35){$u_k-u_L$}

\put(1,10){$\lambda + u_k-u_1$}
\put(83,10){$\cdots$}
\put(126,10){$\lambda +u_k-u_k$}
\put(207,10){$\cdots$}
\put(251,10){$\lambda +u_k§-u_L$}

\put(-2,-10){$c_0$}
\put(47,-10){$c_1$}
\put(297,-10){$c_L = c_0$}
\put(247,-10){$c_{L-1}$}
\put(122,-10){$c_{k-1}$}
\put(172,-10){$c_k$}

\put(-12,23){$b_0$}
\put(52,28){$b_1$}
\put(105,28){$b_{k-1}$}
\put(177,28){$b_{k}$}
\put(230,28){$b_{L-1}$}
\put(304,23){$b_L = b_0$}

\put(-2,55){$a_0$}
\put(47,55){$a_1$}
\put(122,55){$a_{k-1}$}
\put(172,55){$a_k$}
\put(247,55){$a_{L-1}$}
\put(297,55){$a_L=a_0$}

\put(0,25){\circle*{3}}
\put(50,25){\circle*{3}}
\put(125,25){\circle*{3}}
\put(175,25){\circle*{3}}
\put(250,25){\circle*{3}}
\put(300,25){\circle*{3}}

\end{picture}
\\
& \begin{picture}(450,80)(-27,-20)

\put(-25,24){$=$}

\footnotesize
\put(0,0){\line(1,0){125}}
\put(175,0){\line(1,0){175}}
\put(0,0){\line(0,1){50}}
\put(0,25){\line(1,0){125}}
\put(175,25){\line(1,0){175}}
\put(300,0){\line(0,1){50}}
\put(50,0){\line(0,1){50}}
\put(225,0){\line(0,1){50}}
\put(125,0){\line(0,1){50}}
\put(175,0){\line(0,1){50}}
\put(0,50){\line(1,0){125}}
\put(175,50){\line(1,0){175}}
\put(350,0){\line(0,1){50}}

\put(13,35){$u_k-u_1$}
\put(83,35){$\cdots$}
\put(177,10){\tiny{$\lambda + u_k - u_{k+1}$}}
\put(257,35){$\cdots$}
\put(313,35){$u_k-u_L$}

\put(1,10){$\lambda + u_k-u_1$}
\put(83,10){$\cdots$}
\put(180,35){$u_k-u_{k+1}$}
\put(257,10){$\cdots$}
\put(301,10){$\lambda +u_k§-u_L$}

\put(-2,55){$a_0$}
\put(47,55){$a_1$}
\put(122,55){$a_{k-1}$}
\put(172,55){$a_k$}
\put(222,55){$a_{k+1}$}
\put(297,55){$a_{L-1}$}
\put(347,55){$a_L = a_0$}

\put(-12,23){$b_0$}
\put(52,28){$b_1$}
\put(105,28){$b_{k-1}$}
\put(164,28){$b_{k}$}
\put(227,28){$b_{k+1}$}
\put(280,28){$b_{L-1}$}
\put(354,23){$b_L = b_0$}

\put(-2,-10){$c_0$}
\put(47,-10){$c_1$}
\put(347,-10){$c_L = c_0$}
\put(297,-10){$c_{L-1}$}
\put(122,-10){$c_{k-1}$}
\put(172,-10){$c_k$}
\put(222,-10){$c_{k+1}$}

\put(0,25){\circle*{3}}
\put(50,25){\circle*{3}}
\put(125,25){\circle*{3}}
\put(175,25){\circle*{3}}
\put(250,25){\circle*{3}}
\put(300,25){\circle*{3}}

\linethickness{.3mm}

\multiput(125,25)(4,2){13}{\line(0,1){1}}
\multiput(125,25)(4,-2){13}{\line(0,1){1}}
\end{picture}
\end{aligned}
\end{equation*}
\begin{equation*}
\begin{aligned}
& = \rho(u_{k}-u_{k+1})\rho(u_{k+1}-u_k) \times \\
&\quad\begin{picture}(450,65)(-27,-5)

\footnotesize

\put(-25,23){$\times$}

\put(0,0){\line(1,0){125}}
\put(175,0){\line(1,0){125}}
\put(0,0){\line(0,1){50}}
\put(0,25){\line(1,0){125}}
\put(175,25){\line(1,0){125}}
\put(300,0){\line(0,1){50}}
\put(50,0){\line(0,1){50}}
\put(250,0){\line(0,1){50}}
\put(125,0){\line(0,1){50}}
\put(175,0){\line(0,1){50}}
\put(0,50){\line(1,0){125}}
\put(175,50){\line(1,0){125}}

\linethickness{0.3mm}

\qbezier[25](175,50,0)(150,25)(175,0)

\put(13,35){$u_k-u_1$}
\put(83,35){$\cdots$}
\put(207,35){$\cdots$}
\put(263,35){$u_k-u_L$}

\put(1,10){$\lambda + u_k-u_1$}
\put(83,10){$\cdots$}
\put(207,10){$\cdots$}
\put(251,10){$\lambda +u_k§-u_L$}

\put(-2,55){$a_0$}
\put(47,55){$a_1$}
\put(112,55){$a_{k-1}$}
\put(145,55){$a_k$}
\put(172,55){$a_{k+1}$}
\put(247,55){$a_{L-1}$}
\put(297,55){$a_L = a_0$}

\put(-12,23){$b_0$}
\put(52,28){$b_1$}
\put(113,28){$a_{k}$}
\put(177,28){$b_{k+1}$}
\put(230,28){$b_{L-1}$}
\put(304,23){$b_L=b_0$}

\put(-2,-10){$c_0$}
\put(47,-10){$c_1$}
\put(297,-10){$c_L = c_0$}
\put(247,-10){$c_{L-1}$}
\put(112,-10){$c_{k-1}$}
\put(145,-10){$c_k$}
\put(172,-10){$c_{k+1}$}

\multiput(150,0)(0,4){13}{\line(0,1){1}}

\put(0,25){\circle*{3}}
\put(50,25){\circle*{3}}
\put(175,25){\circle*{3}}
\put(250,25){\circle*{3}}
\put(300,25){\circle*{3}}

\end{picture}
\\& = \cdots   
 = \Big( \prod_{\ell = 1 }^L 
   \rho(u_k - u_\ell) \, \rho(u_\ell - u_k) \Big) \, 
   \, \delta^{a_1}_{c_1}  \cdots \delta^{a_L}_{c_L} \, . &&
\end{aligned}
\end{equation*}
Here, the first graph represents the product of the two transfer matrices at
the particular values.  As a consequence of the shift points (\ref{Wvalues})
the $k$-th column reduces to $\delta^{a_{k}}_{b_{k-1}}\delta^{b_{k-1}}_{c_{k}}
= \delta^{a_{k}}_{b_{k-1}} \delta^{a_k}_{c_k}$.  Using the unitarity
(\ref{W_Unitarity}) of the Boltzmann weights in the $(k+1)$-st column an
additional Kronecker delta, $\delta^{a_{k+1}}_{c_{k+1}}$, is produced.
Repetitive use of unitarity proves that the product of the transfer matrices
is essentially proportional to the identity operator
\begin{equation}
 \mathbf{T}(u_k) \, \mathbf{T}(\lambda + u_k) 
  = \Big( \prod_{\ell = 1 }^L 
   \rho(u_k - u_\ell) \, \rho(u_\ell - u_k) \Big) \, \mathbf{1} \, ,
   \qquad k = 1, 2, \cdots, L \, .
\label{inv_id_period}
\end{equation}
The number of these inversion identities is equal to the length $L$ of the
model.  However, due to the identity
\begin{equation}
 \prod_{\ell = 1}^L \mathbf{T}(u_\ell) = \prod_{\ell = 1}^L 
 \mathbf{T}(\lambda + u_\ell) = 
 \Big( \prod_{k,\ell = 1}^L \rho(u_k-u_\ell) \Big)
 \mathbf{1} \, ,
\end{equation}
only $L-1$ of Eqs.~(\ref{inv_id_period}) are independent.
It should be noted that the homogeneous limit of the inversion identities 
(\ref{inv_id_period}) coincides with the $u=0$ limit of the inversion 
identities derived in \cite{Pear87a} for homogeneous face models.

\subsection{Generic integrable open boundary conditions}
The commuting double-row transfer matrix for an open boundary IRF model
with inhomogeneities $\{u_\ell\}$ is
given by \cite{BePO96,AhKo96}
\begin{equation}
  \label{tm_open}
  \begin{aligned}
    &  D_{b_{0} \cdots b_{L}}^{a_{0} \cdots a_{L}} (u) = 
    \sum_{c_i} \Bop[a_0][c_0][b_0][\lambda - u] \\
    & \qquad \times  \left[
      \prod_{\ell = 1}^L
      \Wop[a_{\ell-1}][a_{\ell}][c_{\ell-1}][c_{\ell}][\lambda - u - u_\ell] 
      \Wop[c_{\ell-1}][c_{\ell}][b_{\ell-1}][b_{\ell}][u - u_\ell] \right]
    \BopR[a_{L}][c_L][b_L][u] \, 
\\
&\begin{picture}(400,80)(-60,-20)

\put(-45,18){$=$}

\tiny
\footnotesize
\put(0,20){\line(-1,1){20}}
\put(0,20){\line(-1,-1){20}}
\put(-20,0){\line(0,1){40}}

\multiput(-20,0)(4,0){6}{\line(0,1){1}}
\multiput(-20,40)(4,0){6}{\line(0,1){1}}

\put(0,40){\line(1,0){300}}
\put(0,20){\line(1,0){300}}
\put(0,0){\line(1,0){300}}
\put(0,0){\line(0,1){40}}
\put(50,0){\line(0,1){40}}
\put(125,0){\line(0,1){40}}
\put(175,0){\line(0,1){40}}
\put(250,0){\line(0,1){40}}
\put(300,0){\line(0,1){40}}

\multiput(300,0)(4,0){6}{\line(0,1){1}}
\multiput(300,40)(4,0){6}{\line(0,1){1}}

\put(300,20){\line(1,1){20}}
\put(300,20){\line(1,-1){20}}
\put(320,0){\line(0,1){40}}

\put(0,20){\circle*{3}}
\put(50,20){\circle*{3}}
\put(125,20){\circle*{3}}
\put(175,20){\circle*{3}}
\put(250,20){\circle*{3}}
\put(300,20){\circle*{3}}

\put(-24,44){$a_0$}
\put(-4,44){$a_0$}
\put(46,44){$a_1$}
\put(118,44){$a_{k-1}$}
\put(170,44){$a_k$}
\put(244,44){$a_{L-1}$}
\put(295,44){$a_L$}
\put(315,44){$a_L$}

\put(52,24){$c_1$}
\put(108,24){$c_{k-1}$}
\put(177,24){$c_k$}
\put(232,24){$c_{L-1}$}

\put(-24,-10){$b_0$}
\put(-4,-10){$b_0$}
\put(46,-10){$b_1$}
\put(118,-10){$b_{k-1}$}
\put(170,-10){$b_k$}
\put(244,-10){$b_{L-1}$}
\put(295,-10){$b_L$}
\put(315,-10){$b_L$}

\put(-19,19){$\lambda$--$u$}

\put(15,8){$u - u_1$}
\put(84,8){$\cdots$}
\put(140,8){$u - u_k$}
\put(208,8){$\cdots$}
\put(265,8){$u - u_L$}

\put(5,28){$\lambda - u - u_1$}
\put(84,28){$\cdots$}
\put(129,28){$\lambda - u - u_k$}
\put(208,28){$\cdots$}
\put(254,28){$\lambda - u - u_L$}

\put(310,19){$u$}

\end{picture}
\end{aligned}
\end{equation}
By construction $\mathbf{D}(u)$ enjoys crossing symmetry \cite{BePO96}
\begin{equation}
  \label{tm_open_cross}
 \mathbf{D}(u) = \mathbf{D}(\lambda - u) \, .
\end{equation}

The derivation of inversion identities for the case of open boundary
conditions is more subtle than for the periodic chain, since apart from the
crossing symmetry and unitarity relations, one has to use the YBE (\ref{YBE})
and the local relations satisfied by the boundary weights, in particular
boundary inversion (\ref{bound_inv}) and crossing (\ref{bcross}).  Crossing
symmetry (\ref{tm_open_cross}) of the transfer matrix implies $
\mathbf{D}(u)\,\mathbf{D}(\lambda+u) = \mathbf{D}(u)\,\mathbf{D}(-u)$.
Considering the second form at $u=u_k$, for $k=1,\ldots,L$, we find that the
inversion identities satisfied by the double-row transfer matrices of IRF
models are given by
\begin{equation}
  \begin{split}
    D_{b_{0} \cdots b_{L}}^{a_{0} \cdots a_{L}} (u_k) \,
    D_{d_{0} \cdots d_{L}}^{b_{0} \cdots b_{L}} (-u_k) 
    &=\beta_{a_L}(u_k) \, \beta_{a_L}(-u_k)\,
      \beta_{a_0}(u_k) \, \beta_{a_0}(-u_k)\,
      \rho(2u_k - \lambda) \, \rho(-\lambda - 2u_k) \,\times\\
    &\qquad\times \prod_{\substack{\ell = 1 \cr \ell\neq k}}^{L} 
               \rho(u_k - u_\ell) \, \rho(-u_k + u_\ell) \,
               \rho(u_k + u_\ell) \, \rho(- u_k - u_\ell ) \,\times\\
    &\qquad\times
       \,\delta_{d_0}^{a_0} \, \delta_{d_1}^{a_1} \cdots  \delta_{d_L}^{a_L} \, .
\end{split}
\label{inv_id_open}
\end{equation}
Note that while the product $\mathbf{D}(u)\,\mathbf{D}(-u)$ is diagonal, the
entries depend on the boundary spins $a_0$, $a_L$.  A graphical proof of these
identities is given in the appendix.

We emphasize that the inversion identities (\ref{inv_id_period}) and
(\ref{inv_id_open}) are valid for any IRF model with bulk and boundary weights
satisfying the algebraic relations (\ref{YBE})-(\ref{bcross}).

\section{Application to RSOS models}

As an application of the identities derived above we now consider the
critical restricted solid-on-solid (RSOS) models on a square lattice.  The SOS
Boltzmann weights satisfying the Yang-Baxter equation (\ref{YBE}) are
\cite{AnBF84}
\begin{align}
\label{rsos_W}
  \Wop[d][c][a][b][u] &= \delta_{bd}\,
    \sqrt{\frac{[\,a\,][\,c\,]}{[\,b\,][\,d\,]}}\, 
  \rho(u+\lambda) - \delta_{ac}\, \rho(u) \, ,
  \qquad \rho(u) = \frac{\sin(u-\lambda)}{\sin\lambda} \, ,
\end{align}
with $[x]=\sin(x\lambda) / \sin\lambda$, for height variables satisfying the
SOS condition given in terms of the adjacency matrix $A$ with entries
$A_{ab}=\delta_{a,b+1}+\delta_{a,b-1}$.  The Boltzmann weights satisfy the
unitarity relation (\ref{W_Unitarity}) and crossing symmetry\footnote{%
  Note that the crossing relation in the form (\ref{Wcrossym}) used above can
  be recovered by a gauge transformation of the Boltzmann weights \cite{AhKo96}
  \begin{equation*}
    \Wop[d][c][a][b][u] \to
    \left(\frac{[\,a\,][\,c\,]}{[\,b\,][\,d\,]}\right)^{\frac{u}{2\lambda}}
    \Wop[d][c][a][b][u]\,.
  \end{equation*}
  This gauge does not affect the transfer matrix (\ref{trnsf_mtx_period}) of
  the periodic model (nor, after a similar transformation of the boundary
  weights, that for open boundary conditions, Eq.~(\ref{tm_open})) or the
  corresponding inversion identities.}
\begin{equation}
  \Wop[d][c][a][b][u] = \sqrt{\frac{[\,a\,][\,c\,]}{[\,b\,][\,d\,]}}\,
         \Wop[a][b][d][c][\lambda-u]\,.
\end{equation}
Choosing $\lambda=\pi/r$ and limiting the height variables to take values from
$\mathfrak{S}=\{1,2,\ldots,r-1\}$ subject to the RSOS condition expressed
through the resulting $(r-1)\times(r-1)$ adjacency matrix defines the
restricted SOS model.

\subsection{Periodic boundary conditions}
Considering the RSOS model with inhomogeneities $\{u_k\}_{k=1}^L$ the transfer
matrix is given by Eq. (\ref{trnsf_mtx_period}) and satisfies the $L-1$
independent inversion identities (\ref{inv_id_period}).  The eigenvalues of
$\mathbf{T}(u)$ satisfy a similar identity, i.e.\
\begin{equation}
  \label{rsosp_id}
  \Lambda^{\mathrm{(p)}}(u_k) \Lambda^{\mathrm{(p)}}(\lambda+u_k)
  = 
    \prod_{\ell=1}^L 
    \rho(u_k-u_\ell)\,\rho(u_\ell-u_k)
  \,,
  \quad k=1,\ldots,L-1\,.
\end{equation}
A similar set of equations has recently been obtained for the SOS model with
antiperiodically twisted boundary conditions (corresponding to $a_L=r-a_0$ in
the present context) model by extending Sklyanin's separation of variables
method \cite{Skly92} to the corresponding dynamical six-vertex model
\cite{Nicc13}.  In this approach the RHS of the inversion identity is related
to the quantum determinant of the dynamical vertex mdoel.

Using (\ref{inv_id_period}) or (\ref{rsosp_id}) together with some information
on the analytical properties of the transfer matrix the solution of the
spectral problem is possible: from (\ref{rsos_W}) we find that the transfer
matrix is periodic in $u$ with period $\pi$ (for even length lattices) and
both $\mathbf{T}(u)$ and its eigenvalues can be written as Fourier polynomials
\begin{align}
  \label{rsosp_poly}
  \mathbf{T}(u) = \sum_{n=-L/2}^{L/2} \mathbf{T}_n\,\mathrm{e}^{i2nu}\,.
\end{align}
The spectrum of the transfer matrix can be classified into $r-1$ topological
sectors through the asymptotic behaviour of the transfer matrix eigenvalues,
i.e.\
\begin{align}
\label{rsosp_asy}
  \left(\prod_{\ell=1}^L \mathrm{e}^{\pm i\left(u_\ell+\lambda/2\right)}\right)\,
  \Lambda_{\pm L/2}^{(\mathrm{p})} =  
    \frac{\alpha^{(n)}}{\left(2\sin\lambda\right)^L}\,
    ,
\end{align}
where $\alpha^{(n)}$ take values from the spectrum of the adjacency matrix, i.e.
$\alpha^{(n)}\in\mathrm{spec}(A)=\{2\cos(a\lambda)\}_{a=1}^{r-1}$ \cite{KlPe92}.
Given a possible value of $\Lambda_{\pm L/2}^{(n)}$ the inversion identities
(\ref{inv_id_period}) constitute a system of quadratic equations for the
remaining Fourier coefficients.
For small systems we have verified that they have $2^{(L-1)}$ independent
solutions for each topological sector.  This is more than the $\left({L \atop
    {L/2}}\right)$ corresponding eigenvalues of the transfer matrix of the
\emph{unrestricted} solid-on-solid model.
The RSOS spectrum is known to be the subset of the SOS one with eigenstates of
non-zero norm on the restricted Hilbert space, i.e.\ $a\in\{1,2,\ldots,r-1\}$
\cite{FeVa99}.  We shall return to the identification of the RSOS spectrum
among the solutions of the inversion identities below.

For an efficient computation of transfer matrix eigenvalues for large systems
the identities (\ref{inv_id_period}), however, are not suitable.  For a
generic choice of the inhomogeneities, in particular $u_\ell\ne u_k+\lambda$
for $\ell\ne k$, however, they are formally equivalent to Baxter's
$TQ$-equation
\begin{align}
\label{genTQp}
  \Lambda^{\mathrm{(p)}}(u) \, q(u) = a(u) \, q(u-\lambda) + d(u) \,
  q(u+\lambda) \, , 
\end{align}
restricted to the discrete set of points $u\in\{u_k,u_k+\lambda\}_{k=1}^L$
provided that $a(u_k)=0=d(\lambda+u_k)$ and, as a consequence of
(\ref{rsosp_id}),
\begin{align}
\label{qdetp}
  a(\lambda+u_k) \, d(u_k) = \prod_{\ell=1}^L \rho(u_k-u_\ell) \,
  \rho(u_\ell-u_k)\,. 
\end{align}
In the context of Sklyanin's separation of variables this amounts to a choice
of $a(u)$, $d(u)$ factorizing the quantum determinant of a vertex model
\cite{Skly92}.

$TQ$-equations such as (\ref{genTQp}) holding for arbitrary $u$ are obtained
in the Bethe ansatz formulations of the spectral problem of integrable
systems.  Provided that they allow for a sufficiently simple (e.g.\
polynomial) ansatz for the functions $q(u)$ they can be solved using the Bethe
equations for the finitely many zeroes of these functions.

In the present case of the periodic RSOS model we factorize (\ref{qdetp}) as
\begin{align}
  a(u) \equiv \omega \prod_{\ell=1}^L \frac{\sin(u-u_\ell)}{\sin\lambda}\,,
  \quad
  d(u) \equiv \omega^{-1} \prod_{\ell=1}^L
  \frac{\sin(-u+u_\ell+\lambda)}{\sin\lambda}\, ,
\end{align}
and take the Fourier polynomial
\begin{align}
  q(u) = \prod_{j=1}^M \sin(u-\mu_j)\, ,
\end{align}
parametrized by $M$ complex numbers $\mu_j \equiv i\alpha_j+\lambda/2$ as our
ansatz for the $q$-functions.  As a consequence of the analyticity of the
transfer matrix and its eigenvalues the $\alpha_j$ are determined by the Bethe
equations of the inhomogeneous six-vertex model with twisted boundary
conditions
\begin{equation}
  \label{bae_pbc}
  \omega^2 \prod_{\ell=1}^L
     \frac{\sinh\left(\alpha_j+iu_\ell-\frac{i\lambda}{2}\right)}{
           \sinh\left(\alpha_j+iu_\ell+\frac{i\lambda}{2}\right)}
   = - \prod_{k=1}^M
     \frac{\sinh\left(\alpha_j-\alpha_k-i\lambda\right)}{
     \sinh\left(\alpha_j-\alpha_k+i\lambda\right)}\,,
   \quad j=1,\ldots,M\,.
\end{equation}
Here the twist parameter $\omega$ has to be chosen such that the transfer
matrix eigenvalues $\Lambda^{(p)}(u)$ obtained from the $TQ$-equation shows
the asymptotic behaviour (\ref{rsosp_asy}), i.e.\
\begin{align}
  \omega \, \mathrm{e}^{i\left(\frac{L}{2}-M\right)\lambda}
  + \omega^{-1}\,  \mathrm{e}^{-i\left(\frac{L}{2}-M\right)\lambda}
  \equiv \cos{ a\lambda}\,.
\end{align}

Finally, among the solutions to (\ref{bae_pbc}) the ones corresponding to
eigenvalues of the RSOS model have to be selected.  In previous studies of the
RSOS model the set of Bethe equations (\ref{bae_pbc}) in the homogeneous limit
$u_\ell\equiv0$ has been obtained by embedding the RSOS transfer matrix into
the fusion hierarchy of integrable generalizations of the RSOS model
\cite{BaRe89} and using the algebraic Bethe ansatz \cite{FeVa96a,FeVa99}.  In
Ref.~\cite{BaRe89} it has been conjectured, that the spectrum of the RSOS
model is obtained from configurations of $M=L/2$ roots grouped into strings of
length $n=1,\ldots,r-1$ with real centers $\alpha^{(n)}_{j}$ \cite{TaSu72}
\begin{align}
  \alpha^{(n)}_{j,m} = \alpha^{(n)}_{i} + i\lambda\left(\frac{n+1}{2}-m\right)\,,
  \quad m=1,\ldots,n\,.
\end{align}
Based on this conjecture, integral equations for the densities of these
strings in the thermodynamic limit have been derived and the low energy
effective field theories describing the critical behaviour of the RSOS models
have been identified \cite{Huse84,BaRe89}.

\subsection{Open diagonal boundary conditions}
As a consequence of the adjacency condition for the RSOS model the most
general boundary matrix has the form \cite{AhKo96}
\begin{align}
  \label{bweight}
  \Bop[a][c][b][u] &= \delta_{a\ne b}X_{ab}^c(u) +
  \delta_{a,b}\left(\delta_{c,a+1}D_c(u) + 
    \delta_{c,a-1}U_c(u)\right) \,.
\end{align}
For diagonal boundaries (also named fixed boundaries \cite{BePe96})
the Hilbert space of the model can be decomposed into sectors labelled by the
boundary heights $a_0$ and $a_L$, with the allowed combinations of $(a_0,a_L)$
depending on the length $L$ of the system.  In this case the non-vanishing
boundary weights are given by \cite{AhKo96,BePO96}
\begin{equation}
\label{bdiag}
\begin{aligned}
  D_{a+1}(u) &= \sqrt{\frac{[\,a+1\,]}{[\,a\,]}}\,
     \frac{\sin(u+\xi_a)\sin(u- a\lambda-\xi_a)}{\sin^2\lambda}\,,\\
  U_{a-1}(u) &= \sqrt{\frac{[\,a-1\,]}{[\,a\,]}}\,
     \frac{\sin(u-\xi_a)\sin(u+ a\lambda+\xi_a)}{\sin^2\lambda}\, ,
\end{aligned}
\end{equation}
for $2\le a\le r-2$.  The weights $D_2(u)$ and $U_{r-2}(u)$ simply multiply
the eigenvalues of the transfer matrix in the sectors with $a_{0}\in\{1,r-1\}$,
and similar for $a_L$.  Therefore, any crossing symmetric choice of these
weights satisfies the reflection (\ref{BYBE}) and boundary crossing
symmetry (\ref{bcross}) relations.

The inversion identities of the transfer matrix are given by
Eq.~(\ref{inv_id_open}), with the boundary information being captured by the
functions
\begin{equation}
 \beta_a(u) = \frac{\sin(u - \xi_a)
 \sin(u + a\lambda +  \xi_a)}{\sin^2\lambda}  \, .
 \label{beta_funct}
\end{equation}
In general, the boundary parameters $\xi_a$ are chosen different for the left
and right boundaries. By construction, the double-row transfer matrix of the
RSOS model (and its eigenvalues) is an even Fourier polynomials in $u$
\begin{align}
  \mathbf{D}(u) = \sum_{k=-L-2}^{L+2} \mathbf{D}_k\,\mathrm{e}^{2iku}\,,\quad
  \mathbf{D}_k=\mathbf{D}_{-k}\,\mathrm{e}^{-2ik\lambda}, 
\label{Fourier_exp_tobc}
\end{align}
and becomes diagonal at the special point (note that $D_{b_0\cdots
  b_L}^{a_0\cdots a_L}(u)\propto \delta_{b_0}^{a_0}\delta_{b_L}^{a_L}$ for
diagonal boundaries)
\begin{equation}
  D_{b_0\cdots b_L}^{a_0\cdots a_L}(u=0)
  = 2\cos(\lambda) \, \beta_{a_0}(0) \, \beta_{a_L} (0)
\left(
\prod_{\ell=1}^L \rho(u_\ell)\rho(-u_\ell) \right) \, .\
\label{tdiag_at_zero}
\end{equation}

The corresponding eigenvalues $\Lambda^{(\mathrm{o})}(u)$ will have a Fourier
expansion similar to (\ref{Fourier_exp_tobc}). Taking into account the
crossing symmetry (\ref{tm_open_cross}) inherited from the transfer matrix,
$\Lambda^{(\mathrm{o})}(u) = \Lambda^{(\mathrm{o})}(\lambda -u)$, there are in
total $L+3$ undetermined Fourier coefficients.  The $L$ inversion identities
(\ref{inv_id_open}) and relation (\ref{tdiag_at_zero}) may be supplemented by
the asymptotic behavior of the transfer matrix: analysing the spectra of the
double-row transfer matrix for small lattice lengths $L$, we find that the
leading term is
\begin{equation}
  \Lambda^{(\mathrm{o})}_{L+2} =   \frac{2 \cos (\lambda ) e^{-i \lambda  (L+2)}}
  {(2i \sin \lambda)^{2 L+4}} \, ,
\label{lead_diagb} 
\end{equation}
independent of the boundary spins $a_0$, $a_L$ or of the inhomogeneities of
the lattice.
The subleading Fourier coefficient $\Lambda^{(\mathrm{o})}_{L+1}$, however,
does depend on the choice of the boundary sector $(a_0,a_L)$ but is the same
for all states within this sector.
This fact implies that within a $TQ$-equation formulation, the subleading
order should not depend on the Bethe roots.  This observation together with
the inversion identities (\ref{inv_id_open}) for the eigenvalues and
Eqs.~(\ref{tdiag_at_zero}) and (\ref{lead_diagb}) provides a consistent set of
relations, through which the $L+3$ Fourier coefficients are completely
determined.  We have used this scheme to determine the eigenvalues of the
double-row transfer matrix of RSOS models with $r=4,5,6$ and systems sizes up
to $L=8$.  Comparison with the spectrum obtained by exact diagonalization of
the transfer matrix exhibits perfect agreement.

Having verified, that the identities listed above do in fact capture the full
information required for the computation of the transfer matrix spectrum, we
now formulate a $TQ$-equation which allows for an efficient determination of
the eigenvalues for arbitrary system sizes.  Similar as in the case of
periodic boundary conditions, Eq.~(\ref{genTQp}), we note that the inversion
identities (\ref{inv_id_open}) for the eigenvalues can be interpreted as
conditions for the solvability of the difference equation
\begin{equation}
 \Lambda^{(\mathrm{o})}(u) \, q(u) = a(\lambda -u) \, q(u-\lambda) 
 + a(u) \, q(u+\lambda) \, ,
\label{TQ_diag}
\end{equation}
at the special values $u=u_k$ provided that $a(\lambda - u_k) = a(\lambda +
u_k) = 0$ and 
\begin{equation}
\begin{split}
  a(u_k)a(-u_k) =& \beta_{a_0}(u_k) \beta_{a_0}(-u_k)
                   \beta_{a_L}(u_k) \beta_{a_L}(-u_k)
                   \frac{\rho(-\lambda+2u_k)}{\rho(2u_k)} 
                   \frac{\rho(-\lambda-2u_k)}{\rho(-2u_k)} \cr
  &  \times
  \prod_{j = 1}^L \rho(u_k-u_j)\rho(-u_k-u_j)\rho(-u_k+u_j)\rho(u_k+u_j) \, .
\end{split}
\label{some_cond}
\end{equation}
In (\ref{TQ_diag}) we have used the crossing symmetry (\ref{tm_open_cross}) of
the transfer matrix and assumed that the $q$-function is given by the crossing
symmetric Fourier polynomial
\begin{equation}
 q(u) = \prod_{\ell = 1}^M \sin(u - \mu_\ell) \, \sin(u + \mu_\ell - \lambda) \, ,
\end{equation}
whose degree $M$ is determined below. 

To determine the functions $a(u)$ we require (\ref{some_cond}) to hold for
generic values of the spectral parameter: together with the asymptotic
behavior (\ref{lead_diagb}) of $\Lambda^{(\mathrm{o})}$ this condition is
found to determine uniquely the bulk part (containing factors of $\rho(u)$) of
$a(u)$.  For the factorization of the boundary terms, containing the factors
$\beta_\alpha(u)$, there exist four possible combinations, leading to
different $TQ$-equations.  It is convenient to parametrize these distinct
combinations by introducing the signs $\{\epsilon_0, \epsilon_L\} \in \{\pm 1
\}$.  The generic boundary dependence in $a(u)$ which encompasses all four
possibilities can be written then as $\beta_{a_0}(\epsilon_0 u) \,
\beta_{a_L}(\epsilon_L u)$ and the function $a(u)$ reads
\begin{equation}
 a(u) = \beta_{a_0}(\epsilon_0 u) \, \beta_{a_L}(\epsilon_L u) \,
 \frac{\rho(-\lambda+2u)}{\rho(2u)} 
 \prod_{j=1}^L \rho(u-u_j) \, \rho(u+u_j) \, .
\end{equation}
Comparison with the leading asymptotic behavior (\ref{lead_diagb}) relates 
uniquely the value of $M$ with the boundary heights $a_0, a_L$ for each 
combination
\begin{equation}
M = \frac{L- a_0 \epsilon_0 - a_L\epsilon_L}{2} \, .
\end{equation}
We have numerically verified for small size systems that each of the resulting
$TQ$-equations leads to the correct spectrum of the transfer matrix.
Furthermore, the subleading order coefficient of the Fourier expansion of the
eigenvalues (\ref{TQ_diag}) turns out to be independent of the signs
$\epsilon_0,\epsilon_L$ and of the Bethe roots $\mu_j$
\begin{equation}
  \Lambda_{L+1} =  - \frac{2 e^{i\lambda}\Lambda_{L+2}}{\cos\lambda} 
  \Big( \cos (a_0 \lambda ) \cos (a_0 \lambda +2 \xi_{a_0}) 
  +\cos (a_L \lambda ) \cos (a_L \lambda +2 \text{$\xi_{a_L}$}) 
  + \cos 2 \lambda  
  \sum_{\ell  = 1}^{L} \cos 2\mu_{\ell} \Big) \, ,
  \label{sublead_diagb}
\end{equation}
as expected from our numerical analysis above.  For the particular choice
$\epsilon_0 = - \epsilon_L = - 1$ the $TQ$-equation (\ref{TQ_diag}) coincides
with the one derived for the SOS models via the algebraic Bethe ansatz
starting from a reference state outside the Hilbert space of the RSOS model
\cite{AhKo96}.
Using the analyticity of the transfer matrix one can derive Bethe equations
for the parameters $\mu_j=i\alpha_j+\lambda/2$, $k=1,\ldots,M$, in the
$q$-function for the sector $(a_0,a_L)$:
\begin{equation}
\label{bae_obcd}
\begin{aligned}
  & 
  \prod_{x=0,L}
  \frac{\sinh\left(\alpha_j-i(\epsilon_x\xi_{a_x}-\frac{\lambda}{2})\right)}{
        \sinh\left(\alpha_j+i(\epsilon_x\xi_{a_x}-\frac{\lambda}{2})\right)}\,
  \frac{\sinh\left(\alpha_j
          +i(\epsilon_x(\xi_{a_x}+a_x\lambda)+\frac{\lambda}{2})\right)}{
        \sinh\left(\alpha_j
          -i(\epsilon_x(\xi_{a_x}+a_x\lambda)+\frac{\lambda}{2})\right)}\,
  \times\\
  &\quad\times\prod_{\ell=1}^L
    \frac{\sinh\left(\alpha_j-iu_\ell-\frac{i\lambda}{2}\right)}{
          \sinh\left(\alpha_j-iu_\ell+\frac{i\lambda}{2}\right)}\,
    \frac{\sinh\left(\alpha_j+iu_\ell-\frac{i\lambda}{2}\right)}{
          \sinh\left(\alpha_j+iu_\ell+\frac{i\lambda}{2}\right)}
  \\
  &\qquad\qquad= \prod_{k\ne j}^M 
    \frac{\sinh\left(\alpha_j-\alpha_k-i\lambda\right)}{
          \sinh\left(\alpha_j-\alpha_k+i\lambda\right)}\,
    \frac{\sinh\left(\alpha_j+\alpha_k-i\lambda\right)}{
          \sinh\left(\alpha_j+\alpha_k+i\lambda\right)}\,.
  \end{aligned}
\end{equation}
Here the first line contains the phase shifts associated with reflection from
the left and right boundary, respectively.

\subsection{Open non-diagonal boundaries}
Boundary Boltzmann weights for RSOS models with non-diagonal (or free)
integrable boundaries have been constructed by directly solving the BYBE
\cite{AhKo96,AhYo98} and by using the face-vertex correspondence
\cite{FaHS95}.  In \cite{BePe96}, an alternative realization of non-diagonal
boundary weights was given based on an extension of the diagonal ones
(\ref{bdiag}) with auxiliary face weights.  As a consequence, the spectral
problem of the RSOS models subject to non-diagonal boundary conditions can be
mapped to that with diagonal ones presented in the previous section.

Following Ref.~\cite{BePe96}, the construction is based on the observation
that if $B$ satisfies the BYBE (\ref{BYBE}) then the dressed boundary weight
$B'$ defined by
\begin{equation}
\label{bextd}
\begin{aligned}
 \BopP[a_0][c_0][b_0][u] =& \Bop[a_{-n}][c_{-n}][b_{-n}][u] \, \times\\
 & \times
 \sum_{c_{-1} \cdots c_{-n}} \left[ 
   \prod_{j={-n+1}}^0 
   \Wop[a_{j-1}][a_j][c_{j-1}][c_j][u - u_j]
   \Wop[c_{j-1}][c_j][b_{j-1}][b_j][\lambda-u - u_j]
 \right]\,,
\end{aligned}
\end{equation}
also solves the BYBE.  Here, the lattice has formally been extended by $n$
faces with additional inhomogeneities $\{u_j\}_{j=-(n-1)}^{0}$.  For a
particular choice of the inhomogeneities the additional spin variables on
sites $-n,\ldots,-1$ can be eliminated using fusion projection operators on
(\ref{bextd}) \cite{BePe96}.\footnote{A similar construction allows to
  construct dynamical (operator-valued) boundary matrices for vertex models by
  projection onto subspaces of the additional quantum spaces, see
  \cite{FrSl99,FrPa07b}.}  Note, that as a consequence of the adjacency
condition of the RSOS model all boundary spins $a_0$, $b_0$ have the same
parity.

As an example, general non-diagonal left boundary weights for RSOS models with
even parity boundary spins for $r>4$ can be obtained by starting with the
diagonal ones (\ref{bdiag}) for fixed boundary conditions
$a_{-n}=b_{-n}\equiv\bar{a}=\frac{r}{2}$ ($\frac{r+1}{2}$) for even (odd) $r$,
which are then dressed by $n = \left[\frac{r-1}{2}\right]-1$ auxiliary faces.
Projecting out the auxiliary spins requires to choose the auxiliary
inhomogeneities in (\ref{bextd}) as \cite{BePe96}
\begin{equation}
  \label{bextd_proj}
  u_j = \chi_0+(n+j-1)\lambda\,,\quad j=-(n-1),\ldots,0\,.
\end{equation}
Non-diagonal right boundary weights are constructed in an analogous way.  The
eigenvalues of the RSOS model with such boundary conditions can then be
obtained from the $TQ$-equation Eq.~(\ref{TQ_diag}) for an open RSOS model
with $L+n_L+n_R$ faces.  As a consequence of the constraint (\ref{bextd_proj})
the left boundary phase shift in (\ref{bae_obcd}) are changed to
\begin{equation}
  \label{bphase_extd}
  \begin{aligned}
   &\frac{\sinh\left(\alpha-i(\epsilon_0\xi_{a_0}-\frac{\lambda}{2})\right)}{
        \sinh\left(\alpha+i(\epsilon_0\xi_{a_0}-\frac{\lambda}{2})\right)}\,
    \frac{\sinh\left(\alpha
          +i(\epsilon_0(\xi_{a_0}+a_0\lambda)+\frac{\lambda}{2})\right)}{
        \sinh\left(\alpha
          -i(\epsilon_0(\xi_{a_0}+a_0\lambda)+\frac{\lambda}{2})\right)}\,\times
   \\
   &\quad\times
    \frac{\sinh\left(\alpha+i\chi_0-\frac{i\lambda}{2}\right)}{
          \sinh\left(\alpha-i\chi_0+\frac{i\lambda}{2}\right)}\,
    \frac{\sinh\left(\alpha+i\chi_0-i(n_L-\frac12)\lambda\right)}{
          \sinh\left(\alpha-i\chi_0+i(n_L-\frac12)\lambda\right)}\,,
  \end{aligned}
\end{equation}
with a similar change for the right one.

\section{Discussion}

In this work we investigated inhomogeneous IRF models with different boundary
conditions and were able to derive exact inversion identities satisfied by the
commuting transfer matrices of these models.  
Since our derivation is based on a generic set of local relations satisfied by
the face and boundary weights, these inversion identities are applicable for a
large class of integrable IRF models.
The identities found here are similar to the ones obtained previously for
vertex models by means of separation of variables \cite{Skly92,FrSW08,Nicc12}
or based on local properties of the vertex weights and reflection matrices
\cite{CYSW13a}.

Focusing then on the critical RSOS models with periodic and open boundary
conditions, we have solved the spectral problem of the models by using the
extracted inversion identities. In each case, the set of inversion identities
once complemented with relations emerging from transfer matrix properties,
such as their asymptotic behavior, forms a sufficient set to determine the
eigenvalues of the latter.
In a further step, the spectral problem has been formulated as
a $TQ$-equation and a parametrization of the eigenvalues in terms of roots to
Bethe equations can be derived.

For the periodic case, our results reproduce those obtained in
Ref.~\cite{BaRe89} by means of functional relations arising from the fusion
hierarchy of transfer matrices.  
For diagonal open boundary conditions we have derived four different
$TQ$-equations, each of them yielding the complete spectrum of the RSOS
transfer matrix.  One of the $TQ$-equations found here coincides with that
for the \emph{unrestricted} SOS model with open diagonal boundary conditions
by means of the algebraic Bethe ansatz \cite{AhKo96}.  We note, however, that
this approach cannot be applied for the RSOS model due to the restriction of
spin values.  Furthermore, the embedding the transfer matrix of the latter
into that of the SOS model has been shown for periodic boundary conditions
only \cite{FeVa99}.
Finally, $TQ$-equations and a Bethe ansatz for RSOS models with general
non-diagonal boundary conditions have been derived from the inversion
identities for extended IRF models with diagonal boundary weights dressed by
auxiliary faces \cite{BePe96}.

These results for RSOS models show that the use of inversion identities
provides a basis for the efficient solution of the spectral problem of IRF
models in cases where other Bethe ansatz type approaches may not work.  We
note that our present analysis focussed on the eigenvalues of the transfer
matrix only.  Eigenvectors (or the computation of matrix elements) have not
been considered yet.  Recently, methods related to the ones employed here have
led to considerable advances in vertex models with non-diagonal boundary
conditions \cite{Nicc12,BeCr13,CYSW14}.  Using face-vertex correspondence and,
as a first step, exploring the formal similarity of our results to those
obtained using separation of variables for dynamical vertex models
\cite{Nicc13} similar results can be expected for IRF models.

Another problem which has been successfully addressed using separation of
variables for vertex models and the related spin-$\frac12$ chains is that of
the completeness of the Bethe ansatz \cite{Skly92,FrSW08,KiMN14}.  As has been
discussed above, it easily seen that the number of solutions to the inversion
identities exceeds the dimension of the Hilbert space for the RSOS models.
This is a consequence of the restriction of spin variables and the constraints
imposed by the adjacency condition.  For the periodic RSOS model physical
states have been associated with certain root patterns to the Bethe equations
(\ref{bae_pbc}) \cite{BaRe89}.  For the RSOS model with open boundary
conditions or more general IRF models further work is necessary for this
classification.

\begin{acknowledgments}
This work has been supported by the Deutsche Forschungsgemeinschaft under
grant no. Fr~737/7.
\end{acknowledgments}

%
\appendix
\section{Graphical proof of the inversion identities (\ref{inv_id_open}) for
  open boundary conditions}
For the double-row transfer matrix of the open boundary IRF model
(\ref{tm_open}) we consider the product
\begin{equation*}
  \begin{aligned}
  &D_{b_{0} \cdots b_{L}}^{a_{0} \cdots a_{L}} (u_k) \,
   D_{d_{0} \cdots d_{L}}^{b_{0} \cdots b_{L}} (-u_k)  =
   \\
&
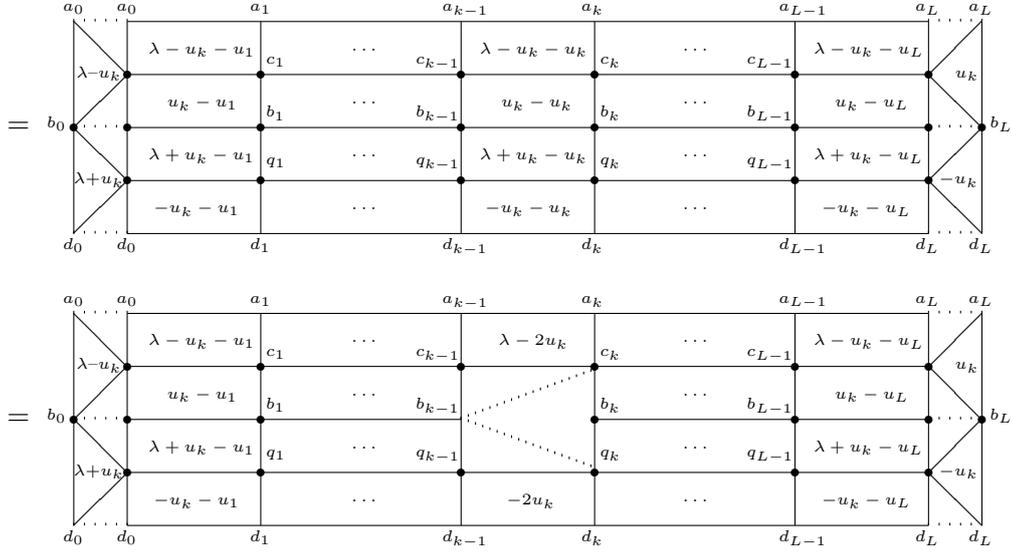
\begin{picture}(450,100)(-80,-50)

\put(-45,-2){$=$}
\tiny

\put(0,20){\line(-1,1){20}}
\put(0,20){\line(-1,-1){20}}
\put(-20,0){\line(0,1){40}}

\multiput(-20,0)(4,0){6}{\line(0,1){1}}
\multiput(-20,40)(4,0){6}{\line(0,1){1}}

\put(0,-20){\line(-1,1){20}}
\put(0,-20){\line(-1,-1){20}}
\put(-20,-40){\line(0,1){40}}

\multiput(-20,-40)(4,0){6}{\line(0,1){1}}

\put(0,40){\line(1,0){300}}
\put(0,20){\line(1,0){300}}
\put(0,0){\line(1,0){300}}
\put(0,0){\line(0,1){40}}
\put(50,0){\line(0,1){40}}
\put(125,0){\line(0,1){40}}
\put(175,0){\line(0,1){40}}
\put(250,0){\line(0,1){40}}
\put(300,0){\line(0,1){40}}

\put(0,-40){\line(1,0){300}}
\put(0,-20){\line(1,0){300}}
\put(0,-40){\line(0,1){40}}
\put(50,-40){\line(0,1){40}}
\put(125,-40){\line(0,1){40}}
\put(175,-40){\line(0,1){40}}
\put(250,-40){\line(0,1){40}}
\put(300,-40){\line(0,1){40}}

\multiput(300,0)(4,0){6}{\line(0,1){1}}
\multiput(300,40)(4,0){6}{\line(0,1){1}}

\put(300,20){\line(1,1){20}}
\put(300,20){\line(1,-1){20}}
\put(320,0){\line(0,1){40}}

\multiput(300,-40)(4,0){6}{\line(0,1){1}}

\put(300,-20){\line(1,1){20}}
\put(300,-20){\line(1,-1){20}}
\put(320,-40){\line(0,1){40}}

\put(0,20){\circle*{3}}
\put(50,20){\circle*{3}}
\put(125,20){\circle*{3}}
\put(175,20){\circle*{3}}
\put(250,20){\circle*{3}}
\put(300,20){\circle*{3}}

\put(0,0){\circle*{3}}
\put(50,0){\circle*{3}}
\put(125,0){\circle*{3}}
\put(175,0){\circle*{3}}
\put(250,0){\circle*{3}}
\put(300,0){\circle*{3}}

\put(0,-20){\circle*{3}}
\put(50,-20){\circle*{3}}
\put(125,-20){\circle*{3}}
\put(175,-20){\circle*{3}}
\put(250,-20){\circle*{3}}
\put(300,-20){\circle*{3}}

\put(-20,0){\circle*{3}}
\put(320,0){\circle*{3}}

\put(-24,44){$a_0$}
\put(-4,44){$a_0$}
\put(46,44){$a_1$}
\put(118,44){$a_{k-1}$}
\put(170,44){$a_k$}
\put(244,44){$a_{L-1}$}
\put(295,44){$a_L$}
\put(315,44){$a_L$}

\put(52,24){$c_1$}
\put(108,24){$c_{k-1}$}
\put(177,24){$c_k$}
\put(232,24){$c_{L-1}$}

\put(52,4){$b_1$}
\put(108,4){$b_{k-1}$}
\put(177,4){$b_k$}
\put(232,4){$b_{L-1}$}

\put(52,-14){$q_1$}
\put(108,-14){$q_{k-1}$}
\put(177,-14){$q_k$}
\put(232,-14){$q_{L-1}$}

\put(-24,-46){$d_0$}
\put(-4,-46){$d_0$}
\put(46,-46){$d_1$}
\put(118,-46){$d_{k-1}$}
\put(170,-46){$d_k$}
\put(244,-46){$d_{L-1}$}
\put(295,-46){$d_L$}
\put(315,-46){$d_L$}

\put(-30,0){$b_0$}
\put(323,0){$b_L$}

\put(-19,19){$\lambda$--$u_k$}
\put(-20,-21){$\lambda$+$u_k$}

\put(8,28){$\lambda - u_k - u_1$}
\put(84,28){$\cdots$}
\put(132,28){$\lambda - u_k - u_k$}
\put(208,28){$\cdots$}
\put(257,28){$\lambda - u_k - u_L$}

\put(15,8){$u_k - u_1$}
\put(84,8){$\cdots$}
\put(140,8){$u_k - u_k$}
\put(208,8){$\cdots$}
\put(265,8){$u_k - u_L$}

\put(8,-12){$\lambda + u_k - u_1$}
\put(84,-12){$\cdots$}
\put(132,-12){$\lambda + u_k - u_k$}
\put(208,-12){$\cdots$}
\put(257,-12){$\lambda + u_k - u_L$}

\put(10,-32){$- u_k - u_1$}
\put(84,-32){$\cdots$}
\put(134,-32){$-u_k - u_k$}
\put(208,-32){$\cdots$}
\put(260,-32){$- u_k - u_L$}

\put(310,19){$u_k$}

\put(304,-21){$-u_k$}

\end{picture}
\\
&\begin{picture}(450,100)(-80,-50)

\put(-45,-2){$=$}
\tiny

\put(0,20){\line(-1,1){20}}
\put(0,20){\line(-1,-1){20}}
\put(-20,0){\line(0,1){40}}

\multiput(-20,0)(4,0){6}{\line(0,1){1}}
\multiput(-20,40)(4,0){6}{\line(0,1){1}}

\put(0,-20){\line(-1,1){20}}
\put(0,-20){\line(-1,-1){20}}
\put(-20,-40){\line(0,1){40}}

\multiput(-20,-40)(4,0){6}{\line(0,1){1}}

\put(0,40){\line(1,0){300}}
\put(0,20){\line(1,0){300}}
\put(0,0){\line(1,0){125}}
\put(175,0){\line(1,0){125}}
\put(0,0){\line(0,1){40}}
\put(50,0){\line(0,1){40}}
\put(125,0){\line(0,1){40}}
\put(175,0){\line(0,1){40}}
\put(250,0){\line(0,1){40}}
\put(300,0){\line(0,1){40}}

\put(0,-40){\line(1,0){300}}
\put(0,-20){\line(1,0){300}}
\put(0,-40){\line(0,1){40}}
\put(50,-40){\line(0,1){40}}
\put(125,-40){\line(0,1){40}}
\put(175,-40){\line(0,1){40}}
\put(250,-40){\line(0,1){40}}
\put(300,-40){\line(0,1){40}}

\multiput(300,0)(4,0){6}{\line(0,1){1}}
\multiput(300,40)(4,0){6}{\line(0,1){1}}

\put(300,20){\line(1,1){20}}
\put(300,20){\line(1,-1){20}}
\put(320,0){\line(0,1){40}}

\multiput(300,-40)(4,0){6}{\line(0,1){1}}

\put(300,-20){\line(1,1){20}}
\put(300,-20){\line(1,-1){20}}
\put(320,-40){\line(0,1){40}}

\put(0,20){\circle*{3}}
\put(50,20){\circle*{3}}
\put(125,20){\circle*{3}}
\put(175,20){\circle*{3}}
\put(250,20){\circle*{3}}
\put(300,20){\circle*{3}}

\put(0,0){\circle*{3}}
\put(50,0){\circle*{3}}
\put(175,0){\circle*{3}}
\put(250,0){\circle*{3}}
\put(300,0){\circle*{3}}

\put(0,-20){\circle*{3}}
\put(50,-20){\circle*{3}}
\put(125,-20){\circle*{3}}
\put(175,-20){\circle*{3}}
\put(250,-20){\circle*{3}}
\put(300,-20){\circle*{3}}

\put(-20,0){\circle*{3}}
\put(320,0){\circle*{3}}

\put(-24,44){$a_0$}
\put(-4,44){$a_0$}
\put(46,44){$a_1$}
\put(118,44){$a_{k-1}$}
\put(170,44){$a_k$}
\put(244,44){$a_{L-1}$}
\put(295,44){$a_L$}
\put(315,44){$a_L$}

\put(52,24){$c_1$}
\put(108,24){$c_{k-1}$}
\put(177,24){$c_k$}
\put(232,24){$c_{L-1}$}

\put(52,4){$b_1$}
\put(108,4){$b_{k-1}$}
\put(177,4){$b_k$}
\put(232,4){$b_{L-1}$}

\put(52,-14){$q_1$}
\put(108,-14){$q_{k-1}$}
\put(177,-14){$q_k$}
\put(232,-14){$q_{L-1}$}

\put(-24,-46){$d_0$}
\put(-4,-46){$d_0$}
\put(46,-46){$d_1$}
\put(118,-46){$d_{k-1}$}
\put(170,-46){$d_k$}
\put(244,-46){$d_{L-1}$}
\put(295,-46){$d_L$}
\put(315,-46){$d_L$}

\put(-30,0){$b_0$}
\put(323,0){$b_L$}

\put(-19,19){$\lambda$--$u_k$}
\put(-20,-21){$\lambda$+$u_k$}

\put(8,28){$\lambda - u_k - u_1$}
\put(84,28){$\cdots$}
\put(139,28){$\lambda - 2u_k$}
\put(208,28){$\cdots$}
\put(257,28){$\lambda - u_k - u_L$}

\put(15,8){$u_k - u_1$}
\put(84,8){$\cdots$}
\put(208,8){$\cdots$}
\put(265,8){$u_k - u_L$}

\put(8,-12){$\lambda + u_k - u_1$}
\put(84,-12){$\cdots$}
\put(208,-12){$\cdots$}
\put(257,-12){$\lambda + u_k - u_L$}

\put(10,-32){$- u_k - u_1$}
\put(84,-32){$\cdots$}
\put(142,-32){$-2u_k$}
\put(208,-32){$\cdots$}
\put(260,-32){$- u_k - u_L$}

\put(310,19){$u_k$}
\put(304,-21){$-u_k$}

\multiput(125,0)(2.7,1){19}{\line(0,1){1}}
\multiput(125,0)(2.7,-1){19}{\line(0,1){1}}

\end{picture}
 \end{aligned}
\end{equation*}
As in the periodic case summations over the spins $b_k$,\ldots,$b_{L-1}$ can
be performed using the unitarity condition (\ref{W_Unitarity}) for the inner
faces resulting in
\begin{equation*}
\begin{aligned}
 & \prod_{\ell =k+1}^L \rho(u_k - u_\ell) \, \rho(u_\ell - u_k) \times 
\\
& 
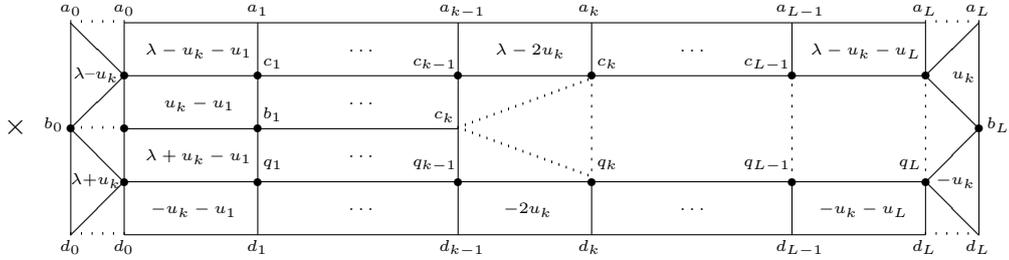
\begin{picture}(450,100)(-80,-50)

\put(-45,-2){$\times$}
\tiny

\put(0,20){\line(-1,1){20}}
\put(0,20){\line(-1,-1){20}}
\put(-20,0){\line(0,1){40}}

\multiput(-20,0)(4,0){6}{\line(0,1){1}}
\multiput(-20,40)(4,0){6}{\line(0,1){1}}

\put(0,-20){\line(-1,1){20}}
\put(0,-20){\line(-1,-1){20}}
\put(-20,-40){\line(0,1){40}}

\multiput(-20,-40)(4,0){6}{\line(0,1){1}}

\put(0,40){\line(1,0){300}}
\put(0,20){\line(1,0){300}}
\put(0,0){\line(1,0){125}}
\put(0,0){\line(0,1){40}}
\put(50,0){\line(0,1){40}}
\put(125,0){\line(0,1){40}}
\put(175,20){\line(0,1){20}}
\put(250,20){\line(0,1){20}}
\put(300,20){\line(0,1){20}}

\put(0,-40){\line(1,0){300}}
\put(0,-20){\line(1,0){300}}
\put(0,-40){\line(0,1){40}}
\put(50,-40){\line(0,1){40}}
\put(125,-40){\line(0,1){40}}
\put(175,-40){\line(0,1){20}}
\put(250,-40){\line(0,1){20}}
\put(300,-40){\line(0,1){20}}

\multiput(300,40)(4,0){6}{\line(0,1){1}}

\put(300,20){\line(1,1){20}}
\put(300,20){\line(1,-1){20}}
\put(320,0){\line(0,1){40}}

\multiput(300,-40)(4,0){6}{\line(0,1){1}}

\put(300,-20){\line(1,1){20}}
\put(300,-20){\line(1,-1){20}}
\put(320,-40){\line(0,1){40}}

\put(0,20){\circle*{3}}
\put(50,20){\circle*{3}}
\put(125,20){\circle*{3}}
\put(175,20){\circle*{3}}
\put(250,20){\circle*{3}}
\put(300,20){\circle*{3}}

\put(0,0){\circle*{3}}
\put(50,0){\circle*{3}}

\put(0,-20){\circle*{3}}
\put(50,-20){\circle*{3}}
\put(125,-20){\circle*{3}}
\put(175,-20){\circle*{3}}
\put(250,-20){\circle*{3}}
\put(300,-20){\circle*{3}}

\put(-20,0){\circle*{3}}
\put(320,0){\circle*{3}}

\put(-24,44){$a_0$}
\put(-4,44){$a_0$}
\put(46,44){$a_1$}
\put(118,44){$a_{k-1}$}
\put(170,44){$a_k$}
\put(244,44){$a_{L-1}$}
\put(295,44){$a_L$}
\put(315,44){$a_L$}

\put(52,24){$c_1$}
\put(108,24){$c_{k-1}$}
\put(177,24){$c_k$}
\put(232,24){$c_{L-1}$}

\put(52,4){$b_1$}
\put(116,4){$c_{k}$}

\put(52,-14){$q_1$}
\put(108,-14){$q_{k-1}$}
\put(177,-14){$q_k$}
\put(232,-14){$q_{L-1}$}
\put(290,-14){$q_{L}$}

\put(-24,-46){$d_0$}
\put(-4,-46){$d_0$}
\put(46,-46){$d_1$}
\put(118,-46){$d_{k-1}$}
\put(170,-46){$d_k$}
\put(244,-46){$d_{L-1}$}
\put(295,-46){$d_L$}
\put(315,-46){$d_L$}

\put(-30,0){$b_0$}
\put(323,0){$b_L$}

\put(-19,19){$\lambda$--$u_k$}
\put(-20,-21){$\lambda$+$u_k$}

\put(8,28){$\lambda - u_k - u_1$}
\put(84,28){$\cdots$}
\put(139,28){$\lambda - 2u_k$}
\put(208,28){$\cdots$}
\put(257,28){$\lambda - u_k - u_L$}

\put(15,8){$u_k - u_1$}
\put(84,8){$\cdots$}

\put(8,-12){$\lambda + u_k - u_1$}
\put(84,-12){$\cdots$}

\put(10,-32){$- u_k - u_1$}
\put(84,-32){$\cdots$}
\put(142,-32){$-2u_k$}
\put(208,-32){$\cdots$}
\put(260,-32){$- u_k - u_L$}

\put(310,19){$u_k$}
\put(304,-21){$-u_k$}

\multiput(175,20)(0,-4){10}{\line(0,1){1}}
\multiput(250,20)(0,-4){10}{\line(0,1){1}}
\multiput(300,20)(0,-4){10}{\line(0,1){1}}

\multiput(125,0)(2.7,1){19}{\line(0,1){1}}
\multiput(125,0)(2.7,-1){19}{\line(0,1){1}}

\normalsize


\end{picture}
\end{aligned}
\end{equation*}
Next, summation over $b_L$ with the boundary inversion condition
(\ref{bound_inv}) gives
\begin{equation*}
\begin{aligned}
& \beta_{a_L}(u_k) \, \beta_{a_L}(-u_k)\,
   \prod_{\ell =k+1}^L \rho(u_k - u_\ell) \, \rho(u_\ell - u_k) \times 
\\
& 
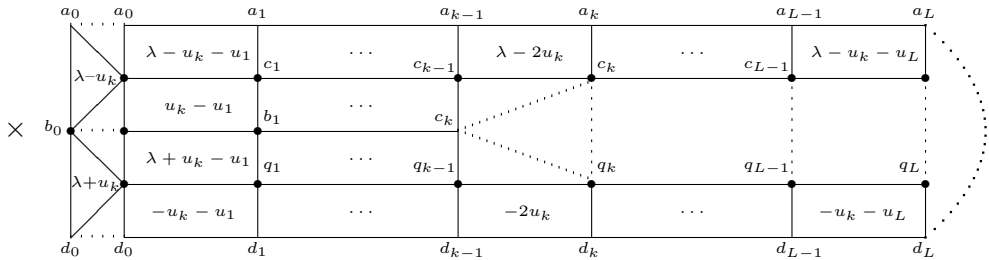
\begin{picture}(450,100)(-80,-50)

\put(-45,-2){$\times$}
\tiny

\put(0,20){\line(-1,1){20}}
\put(0,20){\line(-1,-1){20}}
\put(-20,0){\line(0,1){40}}

\multiput(-20,0)(4,0){6}{\line(0,1){1}}
\multiput(-20,40)(4,0){6}{\line(0,1){1}}

\put(0,-20){\line(-1,1){20}}
\put(0,-20){\line(-1,-1){20}}
\put(-20,-40){\line(0,1){40}}

\multiput(-20,-40)(4,0){6}{\line(0,1){1}}

\put(0,40){\line(1,0){300}}
\put(0,20){\line(1,0){300}}
\put(0,0){\line(1,0){125}}
\put(0,0){\line(0,1){40}}
\put(50,0){\line(0,1){40}}
\put(125,0){\line(0,1){40}}
\put(175,20){\line(0,1){20}}
\put(250,20){\line(0,1){20}}
\put(300,20){\line(0,1){20}}

\put(0,-40){\line(1,0){300}}
\put(0,-20){\line(1,0){300}}
\put(0,-40){\line(0,1){40}}
\put(50,-40){\line(0,1){40}}
\put(125,-40){\line(0,1){40}}
\put(175,-40){\line(0,1){20}}
\put(250,-40){\line(0,1){20}}
\put(300,-40){\line(0,1){20}}

\put(0,20){\circle*{3}}
\put(50,20){\circle*{3}}
\put(125,20){\circle*{3}}
\put(175,20){\circle*{3}}
\put(250,20){\circle*{3}}
\put(300,20){\circle*{3}}

\put(0,0){\circle*{3}}
\put(50,0){\circle*{3}}

\put(0,-20){\circle*{3}}
\put(50,-20){\circle*{3}}
\put(125,-20){\circle*{3}}
\put(175,-20){\circle*{3}}
\put(250,-20){\circle*{3}}
\put(300,-20){\circle*{3}}

\put(-20,0){\circle*{3}}

\put(-24,44){$a_0$}
\put(-4,44){$a_0$}
\put(46,44){$a_1$}
\put(118,44){$a_{k-1}$}
\put(170,44){$a_k$}
\put(244,44){$a_{L-1}$}
\put(295,44){$a_L$}

\put(52,24){$c_1$}
\put(108,24){$c_{k-1}$}
\put(177,24){$c_k$}
\put(232,24){$c_{L-1}$}

\put(52,4){$b_1$}
\put(116,4){$c_{k}$}

\put(52,-14){$q_1$}
\put(108,-14){$q_{k-1}$}
\put(177,-14){$q_k$}
\put(232,-14){$q_{L-1}$}
\put(290,-14){$q_{L}$}

\put(-24,-46){$d_0$}
\put(-4,-46){$d_0$}
\put(46,-46){$d_1$}
\put(118,-46){$d_{k-1}$}
\put(170,-46){$d_k$}
\put(244,-46){$d_{L-1}$}
\put(295,-46){$d_L$}

\put(-30,0){$b_0$}

\put(-19,19){$\lambda$--$u_k$}
\put(-20,-21){$\lambda$+$u_k$}

\put(8,28){$\lambda - u_k - u_1$}
\put(84,28){$\cdots$}
\put(139,28){$\lambda - 2u_k$}
\put(208,28){$\cdots$}
\put(257,28){$\lambda - u_k - u_L$}

\put(15,8){$u_k - u_1$}
\put(84,8){$\cdots$}

\put(8,-12){$\lambda + u_k - u_1$}
\put(84,-12){$\cdots$}

\put(10,-32){$- u_k - u_1$}
\put(84,-32){$\cdots$}
\put(142,-32){$-2u_k$}
\put(208,-32){$\cdots$}
\put(260,-32){$- u_k - u_L$}

\multiput(175,20)(0,-4){10}{\line(0,1){1}}
\multiput(250,20)(0,-4){10}{\line(0,1){1}}
\multiput(300,20)(0,-4){10}{\line(0,1){1}}

\multiput(125,0)(2.7,1){19}{\line(0,1){1}}
\multiput(125,0)(2.7,-1){19}{\line(0,1){1}}

\linethickness{0.3mm}

\qbezier[25](300,40,1)(345,0)(300,-40)

\normalsize


\end{picture}
\end{aligned}
\end{equation*}
Using unitarity again for the outer faces 
we perform the
summation over $q_\ell$, $\ell=L,L-1,\ldots,k+1$
\begin{equation*}
\begin{aligned}
& \beta_{a_L}(u_k) \, \beta_{a_L}(-u_k)\,
    \prod_{\ell =k+1}^L \rho(u_k - u_\ell) \, \rho(-u_k + u_\ell) \,
                        \rho(u_k + u_\ell) \, \rho(- u_k - u_\ell ) \,
    \times \\
&\begin{picture}(450,100)(-80,-50)

\put(-45,-2){$\times$}
\tiny

\put(0,20){\line(-1,1){20}}
\put(0,20){\line(-1,-1){20}}
\put(-20,0){\line(0,1){40}}

\multiput(-20,0)(4,0){6}{\line(0,1){1}}
\multiput(-20,40)(4,0){6}{\line(0,1){1}}

\put(0,-20){\line(-1,1){20}}
\put(0,-20){\line(-1,-1){20}}
\put(-20,-40){\line(0,1){40}}

\multiput(-20,-40)(4,0){6}{\line(0,1){1}}

\put(0,40){\line(1,0){175}}
\put(0,20){\line(1,0){175}}
\put(0,0){\line(1,0){125}}
\put(0,0){\line(0,1){40}}
\put(50,0){\line(0,1){40}}
\put(125,0){\line(0,1){40}}
\put(175,20){\line(0,1){20}}

\put(0,-40){\line(1,0){175}}
\put(0,-20){\line(1,0){175}}
\put(0,-40){\line(0,1){40}}
\put(50,-40){\line(0,1){40}}
\put(125,-40){\line(0,1){40}}
\put(175,-40){\line(0,1){20}}

\put(0,20){\circle*{3}}
\put(50,20){\circle*{3}}
\put(125,20){\circle*{3}}
\put(175,20){\circle*{3}}

\put(0,0){\circle*{3}}
\put(50,0){\circle*{3}}

\put(0,-20){\circle*{3}}
\put(50,-20){\circle*{3}}
\put(125,-20){\circle*{3}}
\put(175,-20){\circle*{3}}

\put(-20,0){\circle*{3}}

\put(-24,44){$a_0$}
\put(-4,44){$a_0$}
\put(46,44){$a_1$}
\put(118,44){$a_{k-1}$}
\put(170,44){$a_k$}

\put(52,24){$c_1$}
\put(108,24){$c_{k-1}$}
\put(177,24){$c_k$}

\put(52,4){$b_1$}
\put(116,4){$c_{k}$}

\put(52,-14){$q_1$}
\put(108,-14){$q_{k-1}$}
\put(177,-25){$q_k$}

\put(-24,-46){$d_0$}
\put(-4,-46){$d_0$}
\put(46,-46){$d_1$}
\put(118,-46){$d_{k-1}$}
\put(170,-46){$d_k$}

\put(-30,0){$b_0$}

\put(-19,19){$\lambda$--$u_k$}
\put(-20,-21){$\lambda$+$u_k$}

\put(8,28){$\lambda - u_k - u_1$}
\put(84,28){$\cdots$}
\put(139,28){$\lambda - 2u_k$}

\put(15,8){$u_k - u_1$}
\put(84,8){$\cdots$}

\put(8,-12){$\lambda + u_k - u_1$}
\put(84,-12){$\cdots$}

\put(10,-32){$- u_k - u_1$}
\put(84,-32){$\cdots$}
\put(142,-32){$-2u_k$}

\multiput(175,20)(0,-4){10}{\line(0,1){1}}

\put(200,44){$a_{k+1}$}
\put(200,-46){$d_{k+1}$}
\put(245,44){$a_{L-1}$}
\put(245,-46){$d_{L-1}$}
\put(297,44){$a_{L}$}
\put(297,-46){$d_{L}$}
\put(220,35){$\cdots$}
\put(220,-35){$\cdots$}

\multiput(125,0)(2.7,1){19}{\line(0,1){1}}
\multiput(125,0)(2.7,-1){19}{\line(0,1){1}}

\linethickness{0.3mm}

\multiput(200,40)(0,-4){20}{\line(0,1){1}}
\multiput(250,40)(0,-4){20}{\line(0,1){1}}
\multiput(300,40)(0,-4){20}{\line(0,1){1}}

\qbezier[25](175,40,1)(220,0)(175,-40)

\normalsize


\end{picture}
\\
& =\,\, \beta_{a_L}(u_k) \, \beta_{a_L}(-u_k)\,
  \prod_{\ell =k+1}^L \rho(u_k - u_\ell) \, \rho(-u_k + u_\ell) \,
  \rho(u_k + u_\ell) \, \rho(- u_k - u_\ell ) \,
  \times \\
& \begin{picture}(450,100)(-80,-50)

\put(-45,-2){$\times$}
\tiny

\put(0,20){\line(-1,1){20}}
\put(0,20){\line(-1,-1){20}}
\put(-20,0){\line(0,1){40}}

\multiput(-20,0)(4,0){6}{\line(0,1){1}}
\multiput(-20,40)(4,0){6}{\line(0,1){1}}

\put(0,-20){\line(-1,1){20}}
\put(0,-20){\line(-1,-1){20}}
\put(-20,-40){\line(0,1){40}}

\multiput(-20,-40)(4,0){6}{\line(0,1){1}}

\put(0,40){\line(1,0){125}}
\put(0,20){\line(1,0){125}}
\put(0,0){\line(1,0){125}}
\put(0,0){\line(0,1){40}}
\put(50,0){\line(0,1){40}}
\put(125,0){\line(0,1){40}}

\put(0,-40){\line(1,0){125}}
\put(0,-20){\line(1,0){125}}
\put(0,-40){\line(0,1){40}}
\put(50,-40){\line(0,1){40}}
\put(125,-40){\line(0,1){40}}

\put(125,20){\line(5,4){25}}
\put(125,20){\line(5,-4){25}}
\put(150,0){\line(5,4){25}}
\put(150,40){\line(5,-4){25}}

\put(125,-20){\line(5,4){25}}
\put(125,-20){\line(5,-4){25}}
\put(150,0){\line(5,-4){25}}
\put(150,-40){\line(5,4){25}}

\multiput(125,40)(4,0){7}{\line(0,1){1}}
\multiput(125,-40)(4,0){7}{\line(0,1){1}}
\multiput(125,0)(4,0){7}{\line(0,1){1}}

\put(150,0){\circle*{3}}

\put(0,20){\circle*{3}}
\put(50,20){\circle*{3}}
\put(125,20){\circle*{3}}

\put(0,0){\circle*{3}}
\put(50,0){\circle*{3}}

\put(0,-20){\circle*{3}}
\put(50,-20){\circle*{3}}
\put(125,-20){\circle*{3}}

\put(-20,0){\circle*{3}}

\put(-24,44){$a_0$}
\put(-4,44){$a_0$}
\put(46,44){$a_1$}
\put(110,44){$a_{k-1}$}
\put(149,44){$a_{k-1}$}

\put(52,24){$c_1$}
\put(108,24){$c_{k-1}$}
\put(177,24){$a_k$}

\put(52,4){$b_1$}
\put(116,4){$c_{k}$}

\put(52,-14){$q_1$}
\put(108,-14){$q_{k-1}$}
\put(177,-25){$d_k$}

\put(-24,-46){$d_0$}
\put(-4,-46){$d_0$}
\put(46,-46){$d_1$}
\put(110,-46){$d_{k-1}$}
\put(152,-46){$d_{k-1}$}

\put(-30,0){$b_0$}

\put(-19,19){$\lambda$--$u_k$}
\put(-20,-21){$\lambda$+$u_k$}

\put(8,28){$\lambda - u_k - u_1$}
\put(84,28){$\cdots$}
\put(139,18){$\lambda - 2u_k$}

\put(15,8){$u_k - u_1$}
\put(84,8){$\cdots$}

\put(8,-12){$\lambda + u_k - u_1$}
\put(84,-12){$\cdots$}

\put(10,-32){$- u_k - u_1$}
\put(84,-32){$\cdots$}
\put(138,-22){$\lambda + 2u_k$}

\multiput(175,20)(0,-4){10}{\line(0,1){1}}

\linethickness{0.3mm}

\multiput(200,40)(0,-4){20}{\line(0,1){1}}
\multiput(250,40)(0,-4){20}{\line(0,1){1}}
\multiput(300,40)(0,-4){20}{\line(0,1){1}}

\put(200,44){$a_{k+1}$}
\put(200,-46){$d_{k+1}$}
\put(245,44){$a_{L-1}$}
\put(245,-46){$d_{L-1}$}
\put(297,44){$a_{L}$}
\put(297,-46){$d_{L}$}
\put(220,35){$\cdots$}
\put(220,-35){$\cdots$}

\end{picture}
\end{aligned}
\end{equation*}
In the last step we have used the crossing symmetry (\ref{bcross}) of the
Boltzmann weights.  
%
Using the YBE (\ref{YBE}) the outer and inner faces can be
interchanged
\begin{equation*}
\begin{aligned}
& \beta_{a_L}(u_k) \, \beta_{a_L}(-u_k)\,
  \prod_{\ell =k+1}^L \rho(u_k - u_\ell) \, \rho(-u_k + u_\ell) \,
  \rho(u_k + u_\ell) \, \rho(- u_k - u_\ell ) \,
  \times \\
& \begin{picture}(450,100)(-80,-50)

\put(-45,-2){$\times$}
\tiny

\put(0,20){\line(-1,1){20}}
\put(0,20){\line(-1,-1){20}}
\put(-20,0){\line(0,1){40}}

\multiput(-20,0)(4,0){18}{\line(0,1){1}}
\multiput(-20,40)(4,0){18}{\line(0,1){1}}

\put(0,-20){\line(-1,1){20}}
\put(0,-20){\line(-1,-1){20}}
\put(-20,-40){\line(0,1){40}}

\multiput(-20,-40)(4,0){18}{\line(0,1){1}}

\put(50,40){\line(1,0){150}}
\put(50,20){\line(1,0){150}}
\put(50,0){\line(1,0){150}}
\put(50,40){\line(0,-1){80}}
\put(150,-40){\line(0,1){80}}
\put(200,-40){\line(0,1){80}}

\put(50,-40){\line(1,0){150}}
\put(50,-20){\line(1,0){150}}
\put(100,-40){\line(0,1){80}}

\put(0,20){\line(5,4){25}}
\put(0,20){\line(5,-4){25}}
\put(25,0){\line(5,4){25}}
\put(25,40){\line(5,-4){25}}

\put(0,-20){\line(5,4){25}}
\put(0,-20){\line(5,-4){25}}
\put(25,0){\line(5,-4){25}}
\put(25,-40){\line(5,4){25}}

\put(12,-22){$\lambda + 2u_k$}
\put(11,18){$\lambda - 2u_k$}

\put(0,20){\circle*{3}}
\put(50,20){\circle*{3}}
\put(100,20){\circle*{3}}
\put(100,0){\circle*{3}}
\put(150,0){\circle*{3}}
\put(150,20){\circle*{3}}
\put(150,-20){\circle*{3}}
\put(25,0){\circle*{3}}
\put(50,0){\circle*{3}}

\put(0,-20){\circle*{3}}
\put(50,-20){\circle*{3}}
\put(100,-20){\circle*{3}}

\put(-20,0){\circle*{3}}
\put(200,0){\circle*{3}}

\put(-24,44){$a_0$}
\put(22,44){$a_0$}
\put(46,44){$a_0$}
\put(97,44){$a_{1}$}
\put(145,44){$a_{k-2}$}
\put(199,44){$a_{k-1}$}

\put(52,23){$c_1$}
\put(202,24){$a_k$}

\put(202,0){$c_k$}

\put(51,-17){$q_1$}

\put(-24,-46){$d_0$}
\put(22,-46){$d_0$}
\put(46,-46){$d_0$}
\put(97,-46){$d_{1}$}
\put(145,-46){$d_{k-2}$}
\put(199,-46){$d_{k-1}$}
\put(202,-27){$d_k$}

\put(-30,0){$b_0$}

\put(-19,19){$\lambda$--$u_k$}
\put(-20,-21){$\lambda$+$u_k$}

\put(56,8){$\lambda - u_k - u_1$}
\put(151,8){$\lambda - u_k - u_{k-1}$}
\put(65,28){$u_k - u_1$}
\put(157,28){$u_k - u_{k-1}$}

\put(120,28){$\cdots$}
\put(120,-12){$\cdots$}
\put(120,8){$\cdots$}
\put(120,-32){$\cdots$}

\put(60,-12){$- u_k - u_1$}
\put(155,-12){$- u_k - u_{k-1}$}
\put(56,-32){$\lambda + u_k - u_1$}
\put(150,-32){$\lambda + u_k - u_{k-1}$}

\linethickness{0.3mm}

\qbezier[25](200,20,0)(230,0)(200,-20)

\multiput(250,40)(0,-4){20}{\line(0,1){1}}
\multiput(300,40)(0,-4){20}{\line(0,1){1}}
\multiput(350,40)(0,-4){20}{\line(0,1){1}}

\put(250,44){$a_{k+1}$}
\put(250,-46){$d_{k+1}$}
\put(295,44){$a_{L-1}$}
\put(295,-46){$d_{L-1}$}
\put(347,44){$a_{L}$}
\put(347,-46){$d_{L}$}
\put(270,35){$\cdots$}
\put(270,-35){$\cdots$}

\end{picture}
\end{aligned}
\end{equation*}
%
and using boundary crossing we obtain
\begin{equation*}
\begin{aligned}
&   \beta_{a_L}(u_k) \, \beta_{a_L}(-u_k)\,
  \rho(2u_k - \lambda) \, \rho(-\lambda - 2u_k) \,
  \prod_{\ell =k+1}^L \rho(u_k - u_\ell) \, \rho(-u_k + u_\ell) \,
  \rho(u_k + u_\ell) \, \rho(- u_k - u_\ell ) \,
  \times  \\
%
& \begin{picture}(450,100)(-80,-50)

\put(-45,-2){$\times$}
\tiny

\put(0,20){\line(-1,1){20}}
\put(0,20){\line(-1,-1){20}}
\put(-20,0){\line(0,1){40}}

\multiput(-20,0)(4,0){6}{\line(0,1){1}}
\multiput(-20,40)(4,0){6}{\line(0,1){1}}

\put(0,-20){\line(-1,1){20}}
\put(0,-20){\line(-1,-1){20}}
\put(-20,-40){\line(0,1){40}}

\multiput(-20,-40)(4,0){6}{\line(0,1){1}}

\put(0,40){\line(1,0){150}}
\put(0,20){\line(1,0){150}}
\put(0,0){\line(1,0){150}}
\put(0,40){\line(0,-1){80}}
\put(100,-40){\line(0,1){80}}
\put(150,-40){\line(0,1){80}}

\put(0,-40){\line(1,0){150}}
\put(0,-20){\line(1,0){150}}
\put(50,-40){\line(0,1){80}}

\put(0,20){\circle*{3}}
\put(0,0){\circle*{3}}
\put(50,20){\circle*{3}}
\put(50,0){\circle*{3}}
\put(100,0){\circle*{3}}
\put(100,20){\circle*{3}}
\put(100,-20){\circle*{3}}
\put(50,0){\circle*{3}}

\put(0,-20){\circle*{3}}
\put(0,-20){\circle*{3}}
\put(50,-20){\circle*{3}}

\put(-20,0){\circle*{3}}
\put(150,0){\circle*{3}}

\put(-24,44){$a_0$}
\put(0,44){$a_0$}
\put(47,44){$a_{1}$}
\put(95,44){$a_{k-2}$}
\put(149,44){$a_{k-1}$}

\put(2,23){$c_1$}
\put(152,24){$a_k$}

\put(152,0){$c_k$}

\put(1,-17){$q_1$}

\put(-24,-46){$d_0$}
\put(0,-46){$d_0$}
\put(47,-46){$d_{1}$}
\put(95,-46){$d_{k-2}$}
\put(149,-46){$d_{k-1}$}
\put(152,-27){$d_k$}

\put(-30,0){$b_0$}
\put(52,4){$b_1$}
\put(84,4){$b_{k-2}$}

\put(-17,19){$u_k$}
\put(-18,-21){$-u_k$}

\put(6,8){$\lambda - u_k - u_1$}
\put(101,8){$\lambda - u_k - u_{k-1}$}
\put(15,28){$u_k - u_1$}
\put(107,28){$u_k - u_{k-1}$}

\put(70,28){$\cdots$}
\put(70,-12){$\cdots$}
\put(70,8){$\cdots$}
\put(70,-32){$\cdots$}

\put(10,-12){$- u_k - u_1$}
\put(105,-12){$- u_k - u_{k-1}$}
\put(6,-32){$\lambda + u_k - u_1$}
\put(101,-32){$\lambda + u_k - u_{k-1}$}

\linethickness{0.3mm}

\qbezier[25](150,20,0)(180,0)(150,-20)

\multiput(200,40)(0,-4){20}{\line(0,1){1}}
\multiput(250,40)(0,-4){20}{\line(0,1){1}}
\multiput(300,40)(0,-4){20}{\line(0,1){1}}

\put(195,44){$a_{k+1}$}
\put(195,-46){$d_{k+1}$}
\put(245,44){$a_{L-1}$}
\put(245,-46){$d_{L-1}$}
\put(297,44){$a_{L}$}
\put(297,-46){$d_{L}$}
\put(220,35){$\cdots$}
\put(220,-35){$\cdots$}

\normalsize


\end{picture}
\end{aligned}
\end{equation*}
Employing unitarity (\ref{W_Unitarity}) for the inner faces the spins $c_k$
and $b_\ell$, $\ell=k-2,\ldots,1$ are summed over
\begin{equation*}
\begin{aligned}
  &\beta_{a_L}(u_k) \, \beta_{a_L}(-u_k)\,
  \rho(2u_k - \lambda) \, \rho(-\lambda - 2u_k) \,\times\\
  &\qquad\times \prod_{\ell =k+1}^L \rho(u_k - u_\ell) \, \rho(-u_k + u_\ell) \,
     \prod_{\substack{\ell = 1 \cr \ell\neq k}}^{L} 
           \rho(u_k + u_\ell) \, \rho(- u_k - u_\ell ) \,
  \times
\\
&
\begin{picture}(450,100)(-80,-50)

\put(-45,-2){$\times$}
\tiny

\put(0,20){\line(-1,1){20}}
\put(0,20){\line(-1,-1){20}}
\put(-20,0){\line(0,1){40}}

\multiput(-20,40)(4,0){6}{\line(0,1){1}}

\put(0,-20){\line(-1,1){20}}
\put(0,-20){\line(-1,-1){20}}
\put(-20,-40){\line(0,1){40}}

\multiput(-20,-40)(4,0){6}{\line(0,1){1}}

\put(0,40){\line(1,0){150}}
\put(0,20){\line(1,0){150}}
\put(0,40){\line(0,-1){20}}
\put(0,-20){\line(0,-1){20}}
\put(100,-40){\line(0,1){20}}
\put(100,20){\line(0,1){20}}
\put(150,-40){\line(0,1){20}}
\put(150,20){\line(0,1){20}}

\put(0,-40){\line(1,0){150}}
\put(0,-20){\line(1,0){150}}
\put(50,-40){\line(0,1){20}}
\put(50,20){\line(0,1){20}}

\put(0,20){\circle*{3}}
\put(50,20){\circle*{3}}
\put(100,20){\circle*{3}}
\put(100,-20){\circle*{3}}
\put(0,-20){\circle*{3}}
\put(0,-20){\circle*{3}}
\put(50,-20){\circle*{3}}
\put(-20,0){\circle*{3}}

\put(-24,44){$a_0$}
\put(0,44){$a_0$}
\put(47,44){$a_{1}$}
\put(95,44){$a_{k-2}$}
\put(149,44){$a_{k-1}$}

\put(2,23){$c_1$}
\put(152,20){$a_k$}

\put(1,-17){$q_1$}

\put(-24,-46){$d_0$}
\put(0,-46){$d_0$}
\put(47,-46){$d_{1}$}
\put(95,-46){$d_{k-2}$}
\put(149,-46){$d_{k-1}$}
\put(152,-22){$d_k$}

\put(-30,0){$b_0$}

\put(-17,19){$u_k$}
\put(-18,-21){$-u_k$}

\put(15,28){$u_k - u_1$}
\put(107,28){$u_k - u_{k-1}$}

\put(70,28){$\cdots$}
\put(70,-32){$\cdots$}

\put(6,-32){$\lambda + u_k - u_1$}
\put(101,-32){$\lambda + u_k - u_{k-1}$}

\multiput(0,20)(0,-4){10}{\line(0,1){1}}
\multiput(50,20)(0,-4){10}{\line(0,1){1}}
\multiput(100,20)(0,-4){10}{\line(0,1){1}}
\multiput(150,20)(0,-4){10}{\line(0,1){1}}

\linethickness{0.3mm}

\multiput(200,40)(0,-4){20}{\line(0,1){1}}
\multiput(250,40)(0,-4){20}{\line(0,1){1}}
\multiput(300,40)(0,-4){20}{\line(0,1){1}}

\put(195,44){$a_{k+1}$}
\put(195,-46){$d_{k+1}$}
\put(245,44){$a_{L-1}$}
\put(245,-46){$d_{L-1}$}
\put(297,44){$a_{L}$}
\put(297,-46){$d_{L}$}
\put(220,35){$\cdots$}
\put(220,-35){$\cdots$}

\normalsize


\end{picture}
\end{aligned}
\end{equation*}
Finally, using the boundary inversion condition (\ref{bound_inv}) and
unitarity of the outer faces the summations over the remaining internal spins
can be done resulting in
\begin{equation*}
\begin{aligned}
  &\beta_{a_L}(u_k) \, \beta_{a_L}(-u_k)\,
   \beta_{a_0}(u_k) \, \beta_{a_0}(-u_k)\,
  \rho(2u_k - \lambda) \, \rho(-\lambda - 2u_k) \,\times\\
  &\qquad\times \prod_{\ell =k+1}^L \rho(u_k - u_\ell) \, \rho(-u_k + u_\ell) \,
  \times \prod_{\substack{\ell = 1 \cr \ell\neq k}}^{L} 
           \rho(u_k + u_\ell) \, \rho(- u_k - u_\ell ) \,
  \times
\\
&\begin{picture}(450,100)(-80,-50)

\put(-45,-2){$\times$}

\tiny

\put(0,40){\line(1,0){150}}
\put(0,20){\line(1,0){150}}
\put(0,40){\line(0,-1){20}}
\put(0,-20){\line(0,-1){20}}
\put(100,-40){\line(0,1){20}}
\put(100,20){\line(0,1){20}}
\put(150,-40){\line(0,1){20}}
\put(150,20){\line(0,1){20}}

\put(0,-40){\line(1,0){150}}
\put(0,-20){\line(1,0){150}}
\put(50,-40){\line(0,1){20}}
\put(50,20){\line(0,1){20}}

\put(0,20){\circle*{3}}
\put(50,20){\circle*{3}}
\put(100,20){\circle*{3}}
\put(100,-20){\circle*{3}}
\put(0,-20){\circle*{3}}
\put(0,-20){\circle*{3}}
\put(50,-20){\circle*{3}}

\put(0,44){$a_0$}
\put(47,44){$a_{1}$}
\put(95,44){$a_{k-2}$}
\put(149,44){$a_{k-1}$}

\put(2,23){$c_1$}
\put(152,20){$a_k$}

\put(1,-17){$q_1$}

\put(0,-46){$d_0$}
\put(47,-46){$d_{1}$}
\put(95,-46){$d_{k-2}$}
\put(149,-46){$d_{k-1}$}
\put(152,-22){$d_k$}

\put(15,28){$u_k - u_1$}
\put(107,28){$u_k - u_{k-1}$}

\put(70,28){$\cdots$}
\put(70,-32){$\cdots$}

\put(6,-32){$\lambda + u_k - u_1$}
\put(101,-32){$\lambda + u_k - u_{k-1}$}

\multiput(0,20)(0,-4){10}{\line(0,1){1}}
\multiput(50,20)(0,-4){10}{\line(0,1){1}}
\multiput(100,20)(0,-4){10}{\line(0,1){1}}
\multiput(150,20)(0,-4){10}{\line(0,1){1}}

\linethickness{0.3mm}

\qbezier[25](0,40,0)(-40,0)(0,-40)

\multiput(200,40)(0,-4){20}{\line(0,1){1}}
\multiput(250,40)(0,-4){20}{\line(0,1){1}}
\multiput(300,40)(0,-4){20}{\line(0,1){1}}

\put(195,44){$a_{k+1}$}
\put(195,-46){$d_{k+1}$}
\put(245,44){$a_{L-1}$}
\put(245,-46){$d_{L-1}$}
\put(297,44){$a_{L}$}
\put(297,-46){$d_{L}$}
\put(220,35){$\cdots$}
\put(220,-35){$\cdots$}

\normalsize


\end{picture}
\\
  &=\beta_{a_L}(u_k) \, \beta_{a_L}(-u_k)\,
   \beta_{a_0}(u_k) \, \beta_{a_0}(-u_k)\,
  \rho(2u_k - \lambda) \, \rho(-\lambda - 2u_k) \,\times\\
  &\qquad\times \prod_{\substack{\ell = 1 \cr \ell\neq k}}^{L} 
     \rho(u_k - u_\ell) \, \rho(-u_k + u_\ell) \,
     \rho(u_k + u_\ell) \, \rho(- u_k - u_\ell ) \,
  \times
  \,\delta_{d_0}^{a_0}
  \, \delta_{d_1}^{a_1} \cdots  \delta_{d_L}^{a_L} \, .
\end{aligned}
\end{equation*}

\end{document}